\documentclass[pre,reprint,superscriptaddress,showpacs,amsmath,amssymb,footinbib]{revtex4}
\usepackage[utf8]{inputenc}
\usepackage{stringenc}
\usepackage{pdfescape}
\usepackage[pdftex]{hyperref}
\usepackage{graphicx}
\usepackage{bm}
\usepackage{blindtext}
\usepackage[caption=false]{subfig}
\usepackage{svg}
\usepackage{placeins}
\usepackage{enumitem}
\usepackage{epstopdf}

\AtBeginDocument{\parindent0em\relax\parsep1ex plus.5ex minus.5ex
    \parskip\parsep}

\newcommand{\derivep}[2]{\frac{\partial #1}{\partial #2}}

\newcommand{\taualpha}[0]{\tau_{\alpha}}
\newcommand{\gammadot}[0]{\dot{\gamma}}
\newcommand{\is}[0]{\mathrm{IS}}
\newcommand{\ms}[0]{\mathrm{MS}}
\newcommand{\mb}[0]{\mathrm{MB}}
\newcommand{\x}[0]{\mathbf{x}}
\newcommand{\xis}[0]{\mathbf{x_{\is}}}

\newcommand{\estart}[0]{\epsilon_{\textrm{start}}}

\newcommand{\epeak}[0]{\epsilon_{\textrm{peak}}}
\newcommand{\eis}[0]{\epsilon_{\is}}
\newcommand{\ems}[0]{\epsilon_{\ms}}
\newcommand{\emb}[0]{\epsilon_{\mb}}

\newcommand{\eispeak}[0]{\epsilon_{\is}^{\textrm{peak}}}

\newcommand{\gammams}[0]{\gamma_{\ms}}

\newcommand{\gammamsone}[0]{\gamma_{\ms}^{1}}

\newcommand{\gammastart}[0]{\gamma_{\textrm{start}}}
\newcommand{\gammaul}[0]{\gamma_{\textrm{ul}}}
\newcommand{\dmin}[0]{D_{\mathrm{min}}}

\newcommand{\sigmaend}[0]{\sigma_{\textrm{end}}}
\newcommand{\sigmanext}[0]{\sigma_{\textrm{next}}}

\newcommand{\gammac}[0]{\gamma_{\text{c,1}}}
 \newcommand{\gammacc}[0]{\gamma_{\text{c,2}}}

\newcommand{\gammanl}[0]{\gamma_{\text{nl}}}

\newcommand{\gammamax}[0]{\gamma_{\textrm{max}}}

\newcommand{\gammapeak}[0]{\gamma_{\textrm{peak}}}

\newcommand{\pcycleone}{p_\textrm{cycle,1}}

\newcommand{\preturn}{\textrm{p}_\textrm{return}}

\makeatletter
\newcommand*{\rom}[1]{\expandafter\@slowromancap\romannumeral #1@}
\makeatother

\newcommand{\internalsupplement}[0]
{%
	\let\oldfootnote\thefootnote%
	\setcounter{footnote}{31336}%
	\footnote{See Supplemental Material at [URL will be inserted by publisher] for more details.}%
	\setcounter{footnote}{\oldfootnote}%
}

\begin{document}

\newcommand{\red}[1]{\textcolor{red}{#1}}
\newcommand{\yellow}[1]{\textcolor{yellow}{#1}}
\newcommand{\green}[1]{\textcolor{green}{#1}}

\title{Shearing small glass-forming systems: a potential energy landscape perspective}

\author{Markus Blank-Burian}
\email[Corresponding author: ]{blankburian@wwu.de}
\affiliation{Institute for Physical Chemistry \\ Westf\"alische Wilhelms-Universit\"at M\"unster, Germany}
\author{Andreas Heuer}
\affiliation{Institute for Physical Chemistry \\ Westf\"alische Wilhelms-Universit\"at M\"unster, Germany}

\begin{abstract}
Understanding the yielding of glass-forming systems upon shearing is notoriously difficult since it is a strong non-equilibrium effect.
Here we show that the concept of the potential energy landscape (PEL), developed for the quiescent state, can be extended to shearing.
When introducing an appropriate coarse graining of the extended PEL for sheared systems, one can distinguish two fundamentally different types of plastic events, namely inherent structures (IS) vs. minimized structure (MS) transitions.
We apply these concepts to non-cyclic and cyclic shearing of small systems, which allow us to characterize the properties of elementary plastic events.
Whereas the general properties of the stress-strain curves are similar to larger systems, a closer analysis reveals significantly different properties.
This allows one to identify the impact of the elastic coupling in larger systems.
The concept of MS enables us to relate the stress overshoot of a single trajectory to the emergence of an MS transition. Furthermore, the occurrence of limit cycles can be characterized in great detail for the small systems and connections to the properties of the PEL can be formulated. Possible implications of our small-system results for the macroscopic limit are discussed.
\end{abstract}

\maketitle

\section{Introduction}
A common feature of amorphous glass forming materials like metallic glasses, colloidal particles or foams is an overshoot in the stress-strain curve when applying a constant shear rate when starting at rest.
During this overshoot period, the material performs a transition from solid-like to fluid-like behavior.
It has been the focus of recent work to study the process of yielding in more detail \cite{Leishangthem2017,Shrivastav2016,Shrivastav2016a,Dubey2016a,Jaiswal2016,Priezjev2017,Koumakis2016,Denisov2015}.
There, we find many different definitions of yield points, each characterizing a different aspect of yielding.

The simplest approach of defining a {\it yield strain} in glasses is taking the strain at the overshoot maximum $\gammapeak$.
For the data of 3D Lennard-Jones-type systems, shown in \cite{Leishangthem2017}, one obtains a value around 0.09, largely independent of system size.
Another approach \cite{Jaiswal2016} uses an overlap function to determine whether there has been a larger rearrangement between two configurations.
The yield is then defined as the point where half of the systems had such a transition.
The same results are obtained by using susceptibilities \cite{Parisi2017}.
When using oscillatory shear also the equality of the storage modulus $G'$ and the loss modulus $G''$ can be used to characterize the transition from elastic to non-elastic behavior \cite{Kawasaki2016}, although yielding a slightly larger value.
This rheological crossover strain can also be related to a sharp transition in the anisotropy of the structure factor \cite{Denisov2015}.

Interestingly, via oscillatory shear experiments and simulations in the low-temperature limit a {\it critical strain} $\gammanl$ can be defined, marking the onset of non-linear behavior \cite{Fiocco2013, Knowlton2014,Regev2015,Kawasaki2016,Leishangthem2017}.
Whereas for strain amplitudes $\gammamax < \gammanl$ one observes limit cycles of the trajectory \cite{Leishangthem2017,Priezjev2017,Kawasaki2016,Fiocco2013}, for $\gammamax > \gammanl$ the system irreversibly leaves its initial region of configuration space.
Whereas $\gammapeak$ only has a minor dependence on system size, the value of $\gammanl$, reported in \cite{Fiocco2013}, decreases from 0.115 to 0.070 when increasing the system size from $N=500$ to $N=32000$.

Already for strain values significantly below $\gammapeak$, plastic events do occur.
This raises the vital question, which plastic events finally give rise to the critical strain and the process of yielding, respectively.
More generally one can ask where irreversibility really starts.
To approach these questions, we will employ a new method of detecting transitions using the framework of the potential energy landscape (PEL).
For this purpose we analyze the properties in the extended PEL $V(\x,\gamma)$, depending on the configuration $\x$ as well as the external strain $\gamma$.
The key idea is to study minima with respect to all $3N+1$ variables which are relevant for the dynamics.
For this purpose we introduce the concept of {\it minimized structures} (MS).
This new partitioning of the extended PEL allows us to shed new light on the process of yielding and the critical strain $\gammanl$, respectively.
As for the quiescent case the PEL approach is most informative for relatively small systems.

The goal of this work is to study the plastic events and in particular the MS transitions of small glass-forming systems (mainly $N = 130$ particles).
They can be regarded as elementary processes of the reorganization processes in glasses under the application of shear.
The ultimate goal is to characterize the different characteristic strain values, described so far, in one general view.

Generally, a quantitative PEL analysis requires the use of small systems \cite{Heuer2008}.
In the standard view, as employed, e.g., in the elasto-plastic models \cite{Martens2012}, the overall response is the effect of local plastic events and the subsequent mutual elastic coupling.
The simulation of small systems allows one to identify the elementary events of the reorganization processes in glasses under the application of shear. Here we mainly use systems with $N=130$ particles. We discuss these events, first, with respect to the impact of MS transitions on properties such as energy and relate these observations to the emergence of the overshoot in the stress-strain curve.
Second, we study oscillatory shear experiments and analyse the possible emergence of limit cycles.
Finally, we attempt to relate the properties of small systems to the phenomena seen for much larger systems.
A similar reasoning has been used before for the quiescent case where the study of small systems together with a model for the elastic coupling could explain several non-trivial features of large systems, such as the time-temperature superposition \cite{Rehwald2012}.

The outline of this work is as follows.
In Section II we start with a summary of technical aspects of our numerical simulations of the model glass-forming system.
In Section III we introduce the MS transitions as a key concept to analyse the dynamics in the PEL upon shearing.
The energetic properties of the overshoot in comparison to relaxation processes in the quiescent system is discussed in Section IV.
In Section V we study the properties of the first MS transition when starting to shear a glass.
The results of the oscillatory shear experiments  are presented in Section VI.
Finally, the results are discussed and summarized in Section VII.
For the sake of clarity we have compiled all the different types of strain values, occurring in this work, in Appendix A.

\section{Technical aspects}
We use a slightly modified form of the standard binary mixture of Lennard-Jones particles \cite{Kob1994} with a cutoff of $r_c = 1.8$ in our 3D molecular dynamics (MD) simulations.
Since the potential energy shows saddle-node bifurcations \cite{Malandro1999}, we took care to make the potential continuous up to the second derivative at the cutoff.
For this purpose we multiply the potential with a polynomial of 5th order in the range $r \in [0.9 r_c, r_c]$ which has no effects on the observables discussed below.
For the shearing we used Lees-Edwards boundary conditions \cite{Lees1972} together with the SLLOD algorithm \cite{Evans1984a}.

For the quiescent case ($\gammadot = 0$) the system mainly resides close to the minima of the PEL for $T \le 0.7$ \cite{Sastry1998}, interrupted by fast jumps between them.
Starting with a configuration at $\gamma = 0$, the associated inherent structure (IS) can be found by minimizing the potential energy $V(\x)$ starting with the current particle positions $\x(t)$ in the trajectory \cite{Stillinger1983}.
Its energy is denoted $\eis$. Furthermore, adjacent IS can be combined into a metabasin (MB) \cite{Stillinger1995e,Doliwa2003c}.
Its energy $\emb$ is characterized by the minimum IS energy within this MB, corresponding to the representative MB configuration.

To prepare starting structures for the shearing simulations deep in the glass, we first performed equilibrium simulation runs at $\gamma = 0$ and $T \ge 0.435$ in order to generate different MB configurations.
For runs constrained to a specific starting MB energy $\estart$, we allowed deviations by at most 0.01.
At $\estart \ge -4.6$ we used at least 128 independent starting configurations.
All other runs used a selection of starting configurations with $\left\langle \estart \right\rangle = -4.6$.
We always mention the average energy per particle.
If not mentioned otherwise, we use $N = 130$ particles for our simulations.

Many dependencies on the starting energy are given in the interval $\estart \in  [-4.60, -4.50]$. Both limits correspond to equilibrium temperatures of $T \approx 0.45$ and $T = 0.7$, respectively, i.e. the regime where the PEL is relevant \cite{Sastry1998}.

\section{PEL under shear}
\begin{figure}[tbp]
    \includegraphics[width=0.45\textwidth]{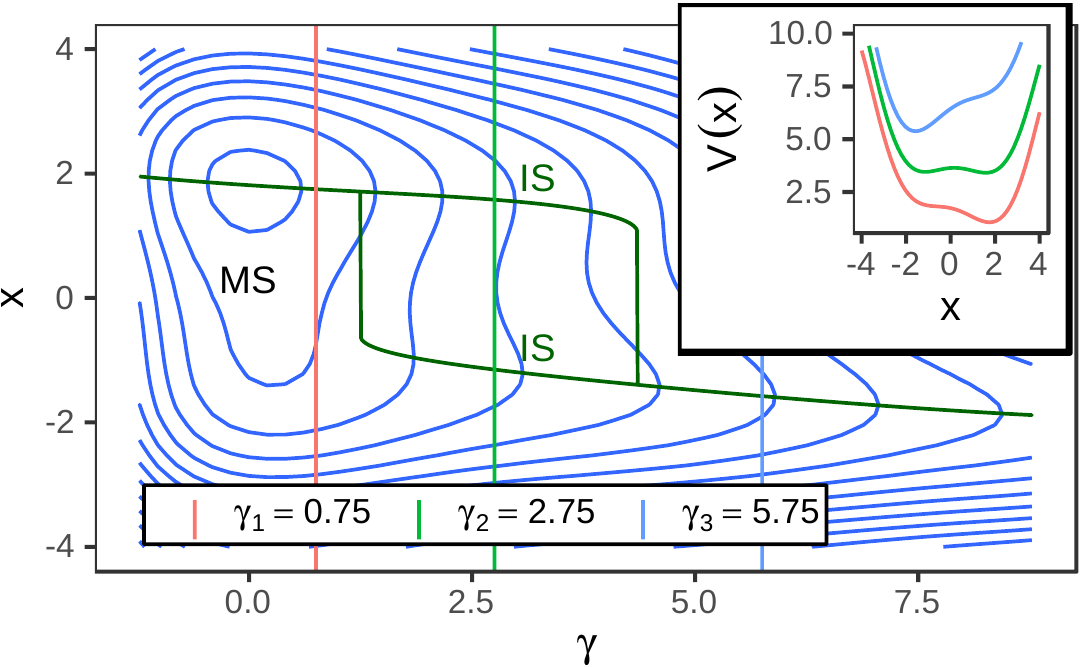}
    \caption{\label{fig:minimization}
    Illustration of the minimization routine for a 1D potential, shown for different strain values.
    Displayed are the contours of a specifically designed 6th order polynomial.
    The solid lines show the taken paths where the inherent structures would evolve by increasing/decreasing the strain.
    The vertical lines mark the strain values, for which the potential is plotted in the inset.
    }
\end{figure}
\begin{figure}[tbp]
    \subfloat[]{
        \label{fig:singleeigenval}
        \includegraphics[width=0.45\textwidth]{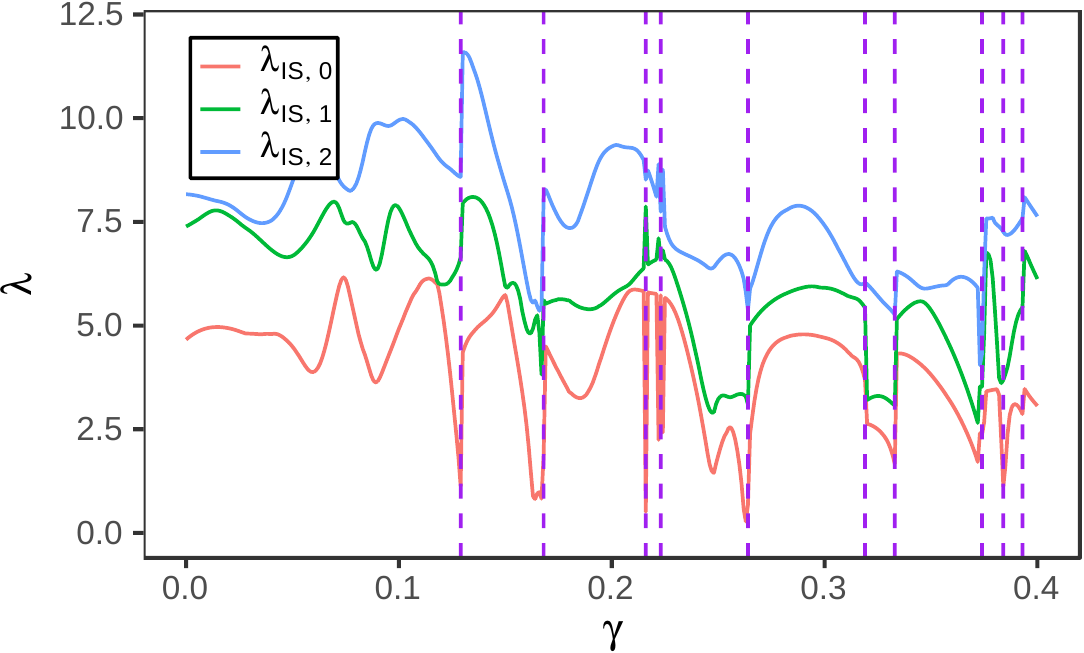}
    }\\
    \subfloat[]{
        \label{fig:singleenergy}
        \includegraphics[width=0.45\textwidth]{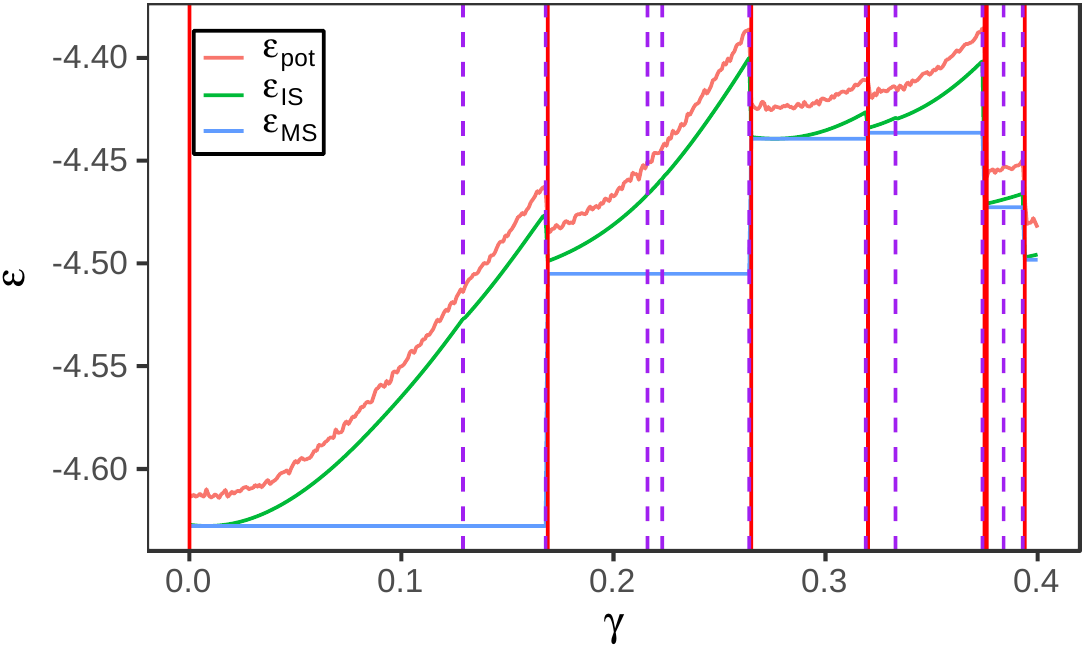}
    }\\
    \subfloat[]{
        \label{fig:singlestress}
        \includegraphics[width=0.45\textwidth]{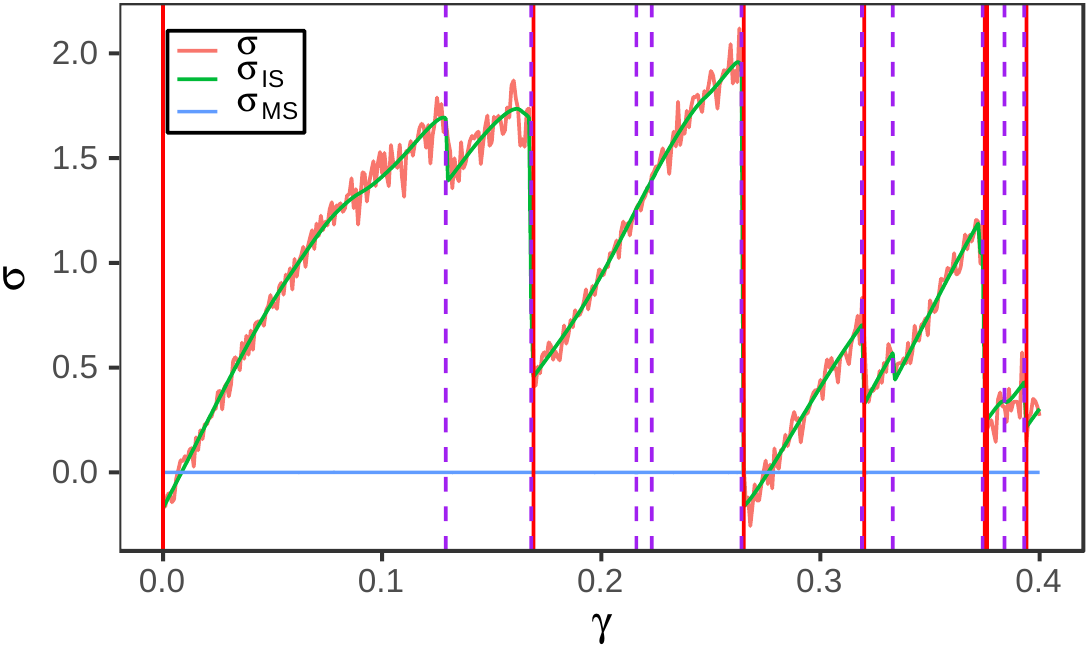}
    }
    \caption{
        \label{fig:single}
        (a) Lowest 3 eigenvalues of the Hessian matrix at the start of a shear simulation, starting at $\epsilon \approx -4.65$ with $T = 0.01$, $N = 130$.
        Dashed/Solid vertical lines mark the detected transitions between inherent structures (IS) / minimized structures (MS).
        (b) Energy and (c) stress of the inherent structures for the same simulation.
    }
\end{figure}
When shearing the system at low temperatures, the configuration closely follows the current local potential energy minimum, defined in the complete potential energy $V(\x,\gamma)$ as also expected for athermal quasistatic shear (AQS) simulations \cite{Maloney2006, Malandro1999}.
For some strain value the barrier disappears and the system experiences  a plastic event, thereby reducing its stress.
On a qualitative level the shearing can be visualized  as a tilting of the PEL \cite{Malandro1999} and a subsequent fold-catastrophe \cite{Chung2012c}.

Our goal is to describe these processes in the extended PEL, based on $V(\x,\gamma)$. The dependence of the potential energy on the strain $\gamma$ results from the use of Lees-Edwards boundary conditions.
For illustration purposes we first discuss a toy PEL $V(\xis,\gamma)$ with just one real-space coordinate $\xis$; see FIG.~\ref{fig:minimization}.
An example trajectory is included which, for a given $\gamma$, is always at the local minimum with respect to the spatial coordinate $\xis$, i.e. in an IS.
As the strain is increased, its coordinate changes and the energy increases.
At some strain value the fold-catastrophe occurs and the system performs a plastic event via a discontinuous change of $\xis$.
This corresponds to an IS transition. In our example, this happens at $\gamma \approx 4.3$.

In addition to the IS we additionally define a minimized structure (MS).
Starting from the present configuration, we minimize $V(\x,\gamma)$ \footnote{
More precise, we first minimize $V(\x, \gamma = \textrm{const})$, which gives us $\xis$.
Then we minimize $V(\xis + S(\gamma) \xis + \Delta \x, \gamma)$ with $\xis = \textrm{const}$ and using the shear matrix $S(\gamma)$.
}.
This is comparable to an AQS simulation along the $\gamma$-direction for which the energy decreases and stopping when the potential energy displays a minimum.
Thus, under cyclic AQS shear, the trajectory would always form a limit-cycle as long as there is no MS transition.
Since the stress at the IS is given by $\sigma = \derivep{}{\gamma}V(\mathbf{x_{\is}}, \gamma)$, the stress at the MS is always zero.
After a jump in the IS trajectory, the new IS may still be connected to the same MS, as exemplified in FIG.~\ref{fig:minimization}.

The same behavior is also shown for an actual trajectory; see FIG.~\ref{fig:single}.
At plastic events, the lowest eigenvalue vanishes due to the fold catastrophes.
In FIG.~\ref{fig:singleeigenval}, this is only partly visible due to a write-step of $\Delta \gamma = 10^{-3}$.
Similarly, the increase of the IS energy $\eis$ as well as the stress is interrupted by drops due to vanishing energy minima.
However, there are several plastic events where the system is still connected to the same MS.

To find transitions in the IS trajectory, we detect peaks of the minimized non-affine $\dmin$ distance \cite{Falk1998} between adjacent points ($\Delta \gamma = 10^{-3}$).
We determined, that using a peak detection algorithm works better than using a simple threshold, as the distribution of energy and stress drops allows for very small values; see Appendix~\ref{app:compisms}, FIG.~\ref{fig:energydrops}.
Comparing the detection rate of $\dmin$ jumps with the same peak detection method applied to the energy or stress, we find that more than 94\% of the detected $\dmin$ jumps are also detected as stress and energy jumps, while the ratio for both other jump types are only 70-80\%.
The exact values can be found in Appendix~\ref{app:jumpdetection}.
In FIG.~\ref{fig:singleeigenval}, we exemplify that our detection method of IS transitions indeed finds the positions, where the eigenvalues drop to zero and are therefore plastic events.

\begin{figure}
    \includegraphics[width=0.45\textwidth]{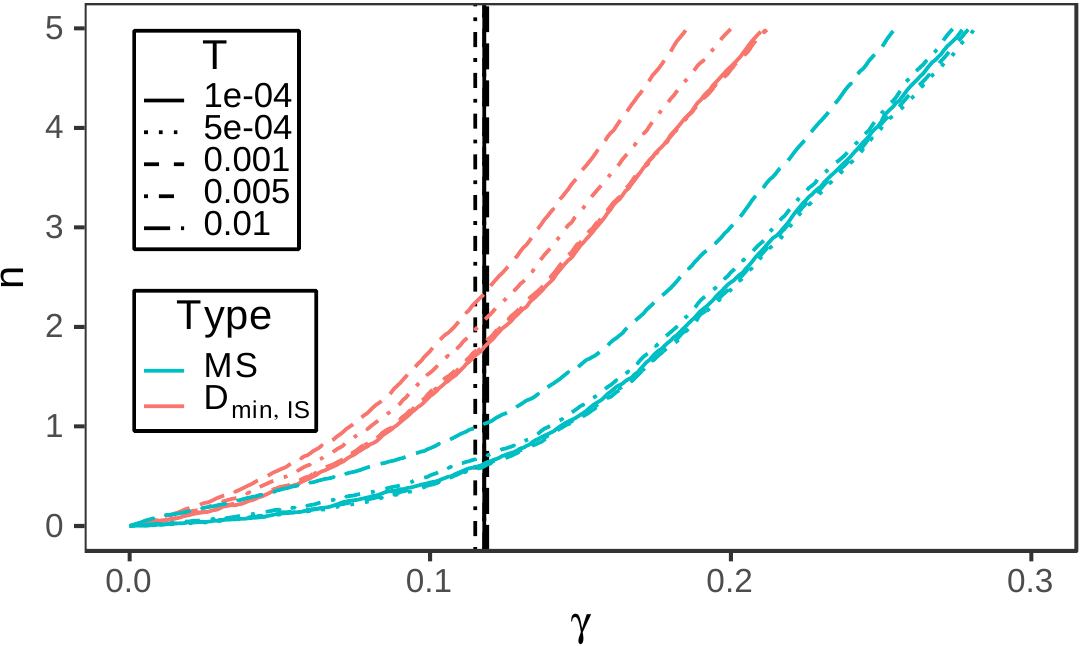}
    \caption{
        Average number of $\ms$ and $\dmin$ transitions since the start of the simulation.
        $\dmin$ transitions are used to detect $\is$ transitions.
        The solid line denotes the strain at the overshoot maximum $\gammapeak$.
    }
    \label{fig:nummb}
\end{figure}

When averaging over many different trajectories, one can determine the average number of MS and IS transitions until some strain $\gamma$; see FIG.~\ref{fig:nummb}.
Note that at the strain of the overshoot maximum on average two plastic events (IS transitions) have occurred but, on average, just half a MS transition.

\section{Relevant barriers for yielding}
\begin{figure}
    \includegraphics[width=0.45\textwidth]{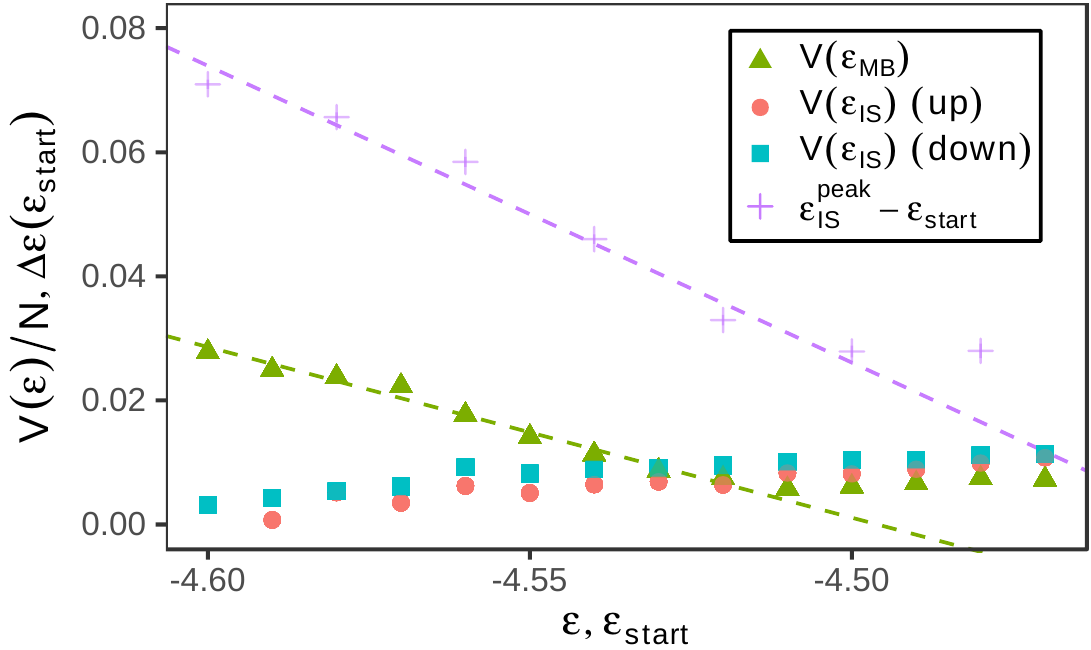}
    \caption{\label{fig:activation}
    IS and MB activation energies for $N = 130$ particles calculated from $\left\langle \tau \middle| \epsilon ; T \right\rangle$.
    The activation energy for jumps to a higher/lower energetic IS is denoted by "up" and "down", respectively.
    We also show the difference between the start energy and energy at the stress peak $\Delta \epsilon = \epeak - \estart$.
    The dashed lines represent linear fits in the range $\emb,\estart \in [-4.6, -4.52]$.
    The slope for $V(\epsilon)/N$ is $-0.275(19)$ and the slope for $\Delta \epsilon$ is $-0.478(52)$.
    }
\end{figure}

As recalled above, in the quiescent case the PEL has two hierarchy levels, namely IS and MB, giving rise to IS-IS and MB-MB transitions at finite temperatures.
The transitions between MB govern the irreversible relaxation of the system above the glass transition.
The typical barriers of IS-IS transitions $V(\eis)$ and of MB-MB transitions $V(\emb)$, respectively, are shown in FIG.~\ref{fig:activation} as a function of the respective energies.
They have been determined by simulations at different temperatures using Arrhenius fits on the IS and MB lifetimes $\langle \tau | \emb ; T\rangle$ \cite{Heuer2008}.
Note that $V(\emb)$ depends linearly on the metabasin energy $\emb$ for $\emb <= -4.52$.
As a consequence, the PEL can be regarded as an assembly of traps with similar absolute barrier energy but different trap depths $\emb$ \cite{Bouchaud2005, Heuer2008}.
In marked contrast, $V(\eis)$ is basically independent of the initial IS-energy.

Naturally, by applying shear to the energy landscape, not only the basins are tilted, but also the whole metabasins.
Thereby, both barriers are changed simultaneously, when applying pure shear to a system.
As metabasins are a superstructure of inherent structures and have much higher barriers at low energies, when starting in the bottom-most IS of a metabasin, we expect some IS transitions (plastic events) before the metabasin barrier vanishes.
Intuitively, one might relate yielding to a crossing of an MB barrier rather than an IS barrier because in analogy to the MB transitions in the quiescent case, yielding reflects an irreversible relaxation process.
To analyze this question, we study how the IS energy at the yielding point $\eispeak$ depends on the initial starting energy $\estart$ by plotting $\eispeak - \estart$ vs. $\estart$; see FIG.~\ref{fig:activation}.
The similarity with the energy dependence of the MB activation energies is striking.
It displays a linear energy dependence for low energies ($\estart \le -4.52$) with similar but not identical slopes and a plateau behavior for higher energies.
Roughly speaking, in the relevant low-energy regime the energy at the yielding point is approx. 0.04 higher than the MB.
Furthermore, in the whole low-energy range the additional modification of the PEL due to shearing is somewhat stronger than typical MB barrier heights.
For the lowest energy -4.60, both energies read 0.04 ($\eispeak - \estart - V(\emb)$) and 0.03 ($V(\emb)$), respectively; see FIG.~\ref{fig:activation}.

In contrast, no similarity exists with the energy dependence of the IS barriers.
These observations suggest, that yielding is indeed related to transitions between the superstructures of the PEL, as captured by the concept of metabasins.
However, a more quantitative analysis proves difficult, as both the barrier and the current IS energy change during shear.

\section{\label{sec:firstms}The first MS transition}
\begin{figure}[tbp]
    \subfloat[Stress]{
        \includegraphics[width=0.5\textwidth]{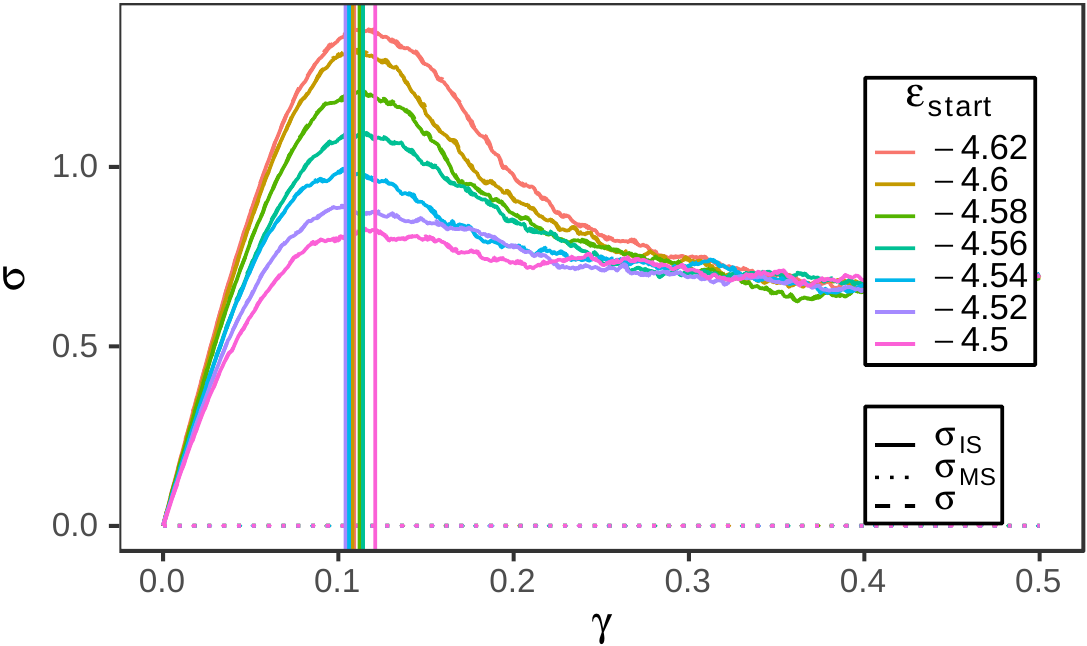}
        \label{fig:stressens}
    }\\
    \subfloat[Energy]{
        \includegraphics[width=0.5\textwidth]{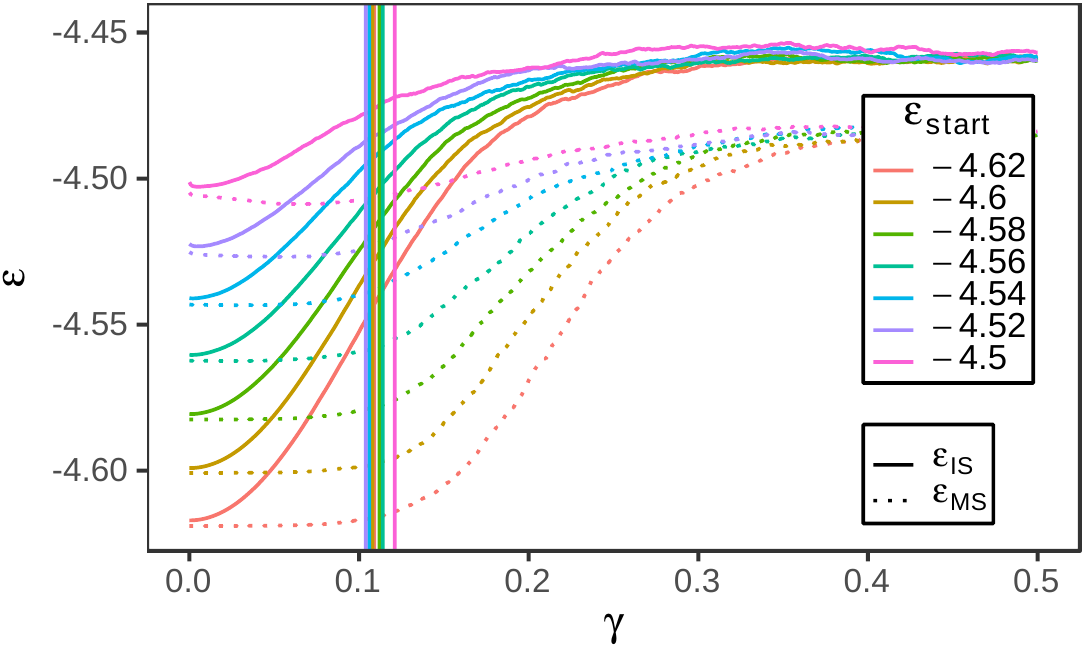}
        \label{fig:energyens}
    }
    \label{fig:ensemble}
    \caption{
        Ensemble averaged (a) energy and (b) stress.
        Solid lines represent IS energies/stresses and dotted lines MS energies/stresses.
        Vertical lines show the position of the maximum of the stress overshoot for the respective starting energy.
    }
\end{figure}

During simulations of start-up shear, we observe that the stress overshoot changes its height in dependence of the starting energy $\estart$, but hardly its position (FIG.~\ref{fig:stressens}, in particular if neglecting the highest starting energy).
Due to simulating at low temperatures, the system stress is close to the IS stress.
The position of the overshoot peak occurs roughly at the same position, where the sign of the curvature in the $\eis$ trajectory changes (FIG.~\ref{fig:energyens}).
However, in the $\ems$ trajectory, we notice one additional feature: The average MS energy does not change significantly up until the position of the overshoot.
As our simulations are starting at the bottom of a metabasin, we expect no transitions to lower energies.
Indeed, the MS energy in single trajectories stays roughly constant in this strain regime but starts to increase afterwards.
This suggests to ask whether the first MS transition is directly related to the stress overshoot.
The strain value, at which in a specific realization the first MS transition occurs, is denoted $\gammamsone$.

\begin{figure}[tbp]
    \includegraphics[width=0.5\textwidth]{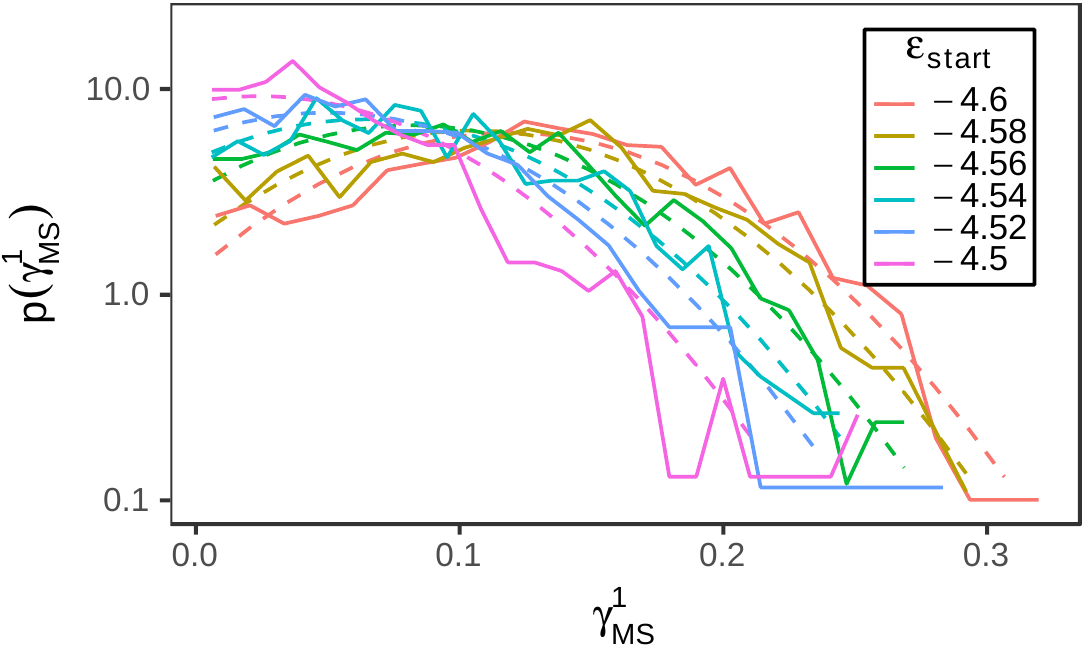}
    \caption{
        Probability distribution of first MS transition $\gammamsone$.
        The dashed lines represent fits with Gaussian distributions having a fixed width $\sigma = 0.0675$ and a cutoff at $\gammamsone = 0$.
    }
    \label{fig:gammams_distribution}
\end{figure}

When averaging over the different trajectories, we can extract the distribution of $\gammamsone$.
It has a Gaussian form with approximately the same width for all $\estart$ and a natural cutoff at $\gamma = 0.0$, as can be seen in FIG.~\ref{fig:gammams_distribution}.
The parameterization of the distribution with respect to $\estart$ can be found in Appendix~\ref{sec:firstms}.
It is remarkable, that while the average position of $\gammamsone$ changes dramatically with $\estart$, the MS energy stays nearly constant until the overshoot for all values of $\estart$ (FIG.~\ref{fig:energyens}).

\begin{figure*}[tbp]
    \subfloat[]{
        \label{fig:enextms}
        \includegraphics[width=0.31\textwidth]{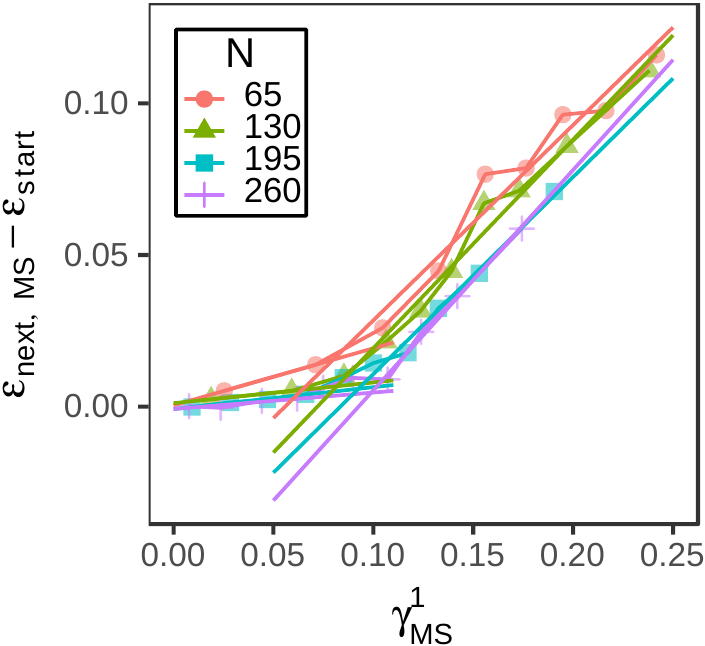}
    }
    \subfloat[]{
        \label{fig:dgamma}
        \includegraphics[width=0.31\textwidth]{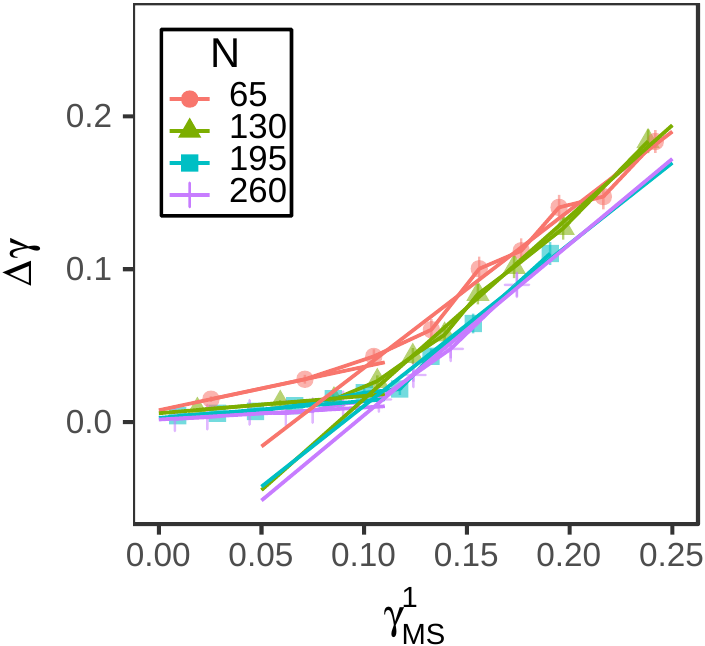}
    }
    \subfloat[]{
        \label{fig:gammmsdjumpmsN}
        \includegraphics[width=0.31\textwidth]{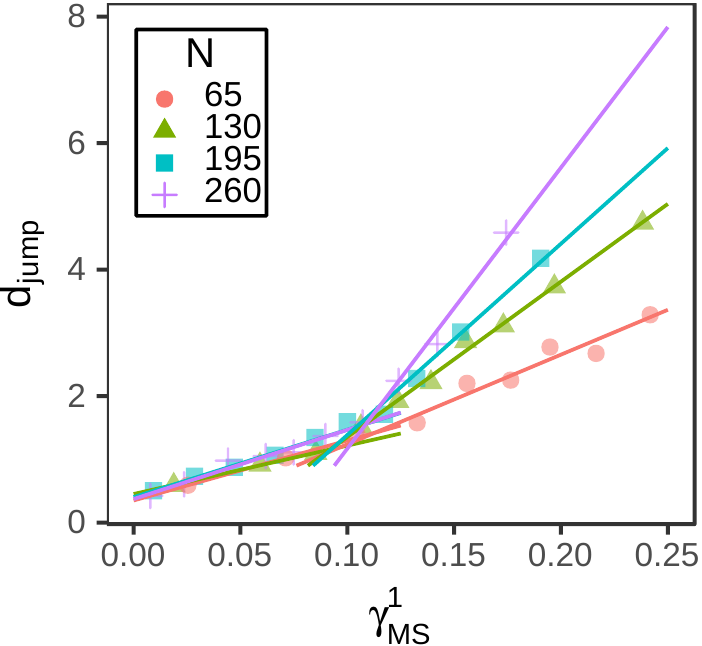}
    }\\
    \caption{\label{fig:gammams}
Change of MS-properties after the first MS-transition as function of $\gammamsone$, i.e. the strain, where this transition occurs.
Included are linear fits for $\gammamsone \le 0.05$ and $\gammamsone \in [0.1,0.2]$.
        (a) Change in MS energy. Intersection point: $\gammamsone = 0.08$ for $N = 130$.
        (b) Change in MS strain. Intersection point:  $\gammamsone = 0.10$ for $N = 130$.
        (c) Euclidean distance between first and second MS. Average intersection point: $\gammamsone = 0.10$.
    }
\end{figure*}

To obtain a closer insight, we analyze the average change in MS energy and in strain as well as the average Euclidean distance between the first and the second MS configuration.
All three observables display two distinct regimes with respect to the value of $\gammamsone$, see FIG.~\ref{fig:gammams}, with a crossover in the strain range around $0.09$ $(\pm 0.01)$.
This crossover strain is abbreviated as $\gammac$.
For an average $\gammamsone < \gammac$, the MS properties hardly change.
In contrast, for $\gammamsone > \gammac$ the dependence on $\gammamsone$ is much stronger.
It turns out (data not shown) that for $T \le 0.01$ no dependence on temperature can be seen.

Additional information can be gained from studying finite-size effects. For the Euclidean distance in FIG.~\ref{fig:gammmsdjumpmsN}.
no finite-size effects are present for $\gammamsone < \gammac$, whereas for  $\gammamsone > \gammac$ the distance increases linearly with increasing system size in the range of studied system sizes. The slope is approximately proportional to the system size.
Thus the first MS transition is very localized for $\gammamsone < \gammac$ and spatially extended otherwise.
Interestingly, for much larger systems an increase of the avalanche size with system size is observed for $\gammamax > \gammanl$ whereas no dependence is observed for $\gammamax < \gammanl$ \cite{Leishangthem2017}. However, in that case the crossover is observed around $\gammanl$ whereas for the small systems, studied in this work, the value of $\gammanl$ (see below) is much larger than $\gammac$. Nevertheless, as argued below, both observations may be directly connected. Furthermore, we also observe in FIG.~\ref{fig:enextms} that the energy per particle during the first MS transition only displays a weak dependence on system size. This implies that the total energy is approximately proportional to the system size. This is compatible with the results for the 
Euclidean distance. Finally, in FIG.~\ref{fig:dgamma} we see that roughly independent of system size the new MS has a strain value which is approximately 0.06 smaller (for $\gammamsone > \gammac$) than the strain value  at which the first MS transition occurs.

\begin{figure}[tbp]
    \includegraphics[width=0.45\textwidth]{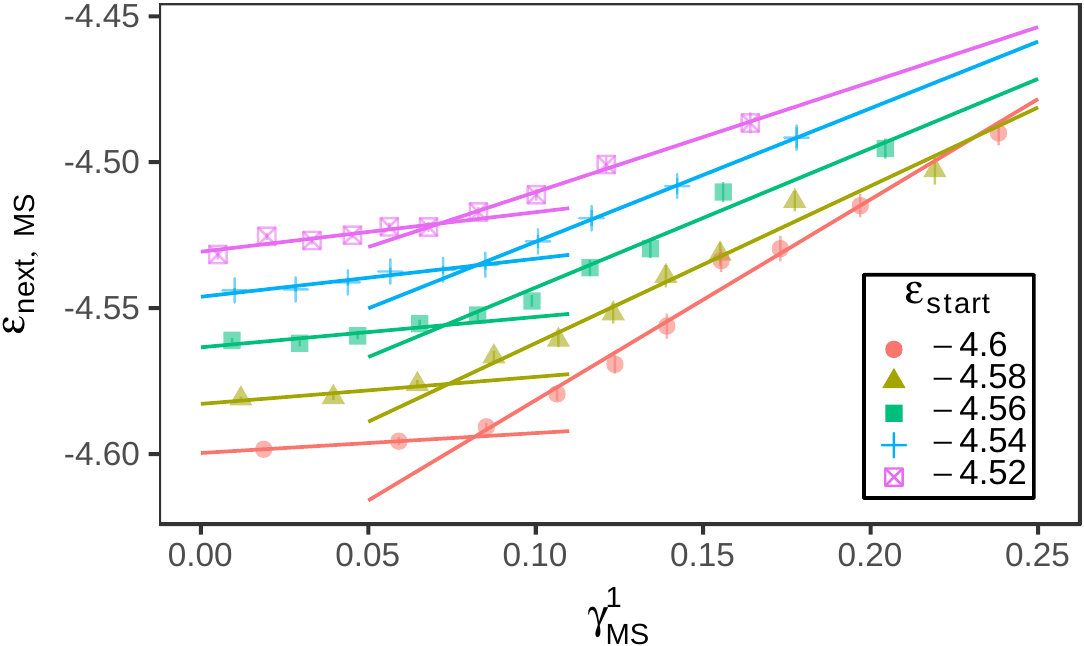}
    \caption{\label{fig:gammams_enextms_IE}%
        MS energy after the first MS transition with respect to the start energy $\estart$.
        The vertical line marks the position of the average crossover at $\gamma = 0.0765(46)$.
    }
\end{figure}

Overall, the results suggest that most MS transitions, occurring at $\gamma < \gammac$, are irrelevant in terms of our analyzed observables.
This highlights the very stable, elastic structure of the glassy system, which is, apart from minor local rearrangements, robust against small perturbations.
In contrast, for $\gamma > \gammac$ the system mainly performs relevant MS transitions.
We remark, that the position of $\gammac$ does not change significantly with the starting energy $\estart$, as can be seen in FIG.~\ref{fig:gammams_enextms_IE}. Thus, the definition of $\gammac$ and thus the presence of relevant and irrelevant MS transitions is basically independent of $\estart$ or, equivalently, the age of the system. Consequences of this observation will be discussed in the final section.

\begin{figure}[tbp]
    \centering
    \includegraphics[width=0.45\textwidth]{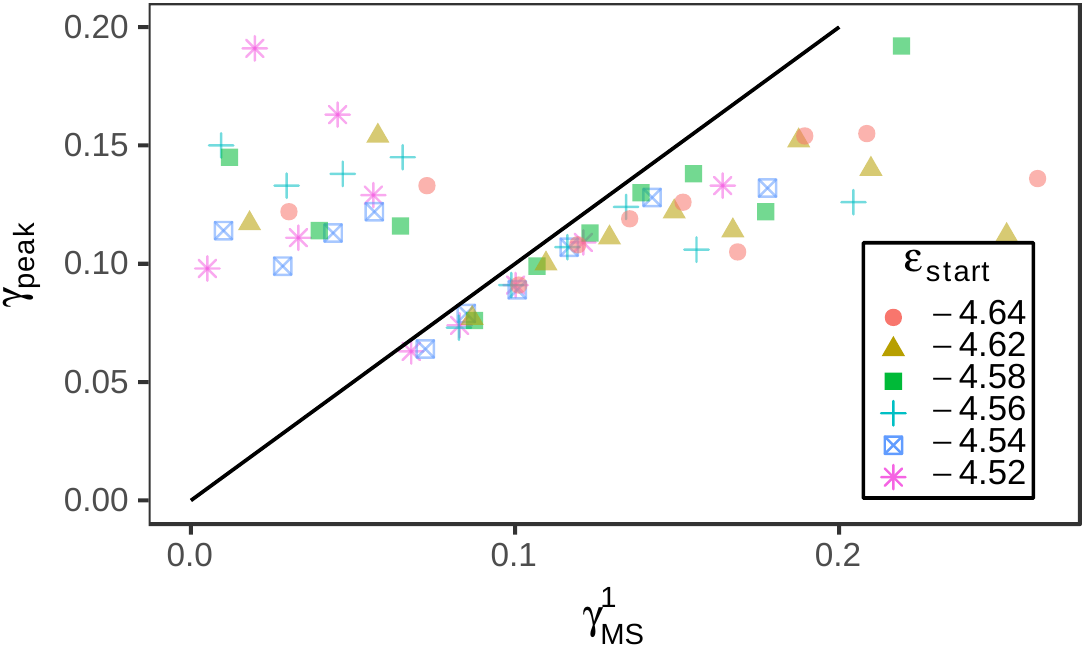}
    \caption{\label{fig:gammams_peak}
    Average stress overshoot peak position for all trajectories with the MS transition at a given strain.
    The solid black line reflects the diagonal.
    }
\end{figure}

We can also gain insight into the position of the stress overshoot maximum $\gammapeak$ and rationalize why $\gammac < \gammapeak$.
After binning the individual stress-strain curves (like the example in FIG.~\ref{fig:singlestress}) for trajectories with similar values of $\gammamsone$ one can determine $\gammapeak$ for this subset of trajectories.
These preaveraged curves can be found in Appendix~\ref{sec:firstms} FIG.~\ref{fig:smsubstress}.
FIG.~\ref{fig:gammams_peak} shows the dependence of $\gammapeak$ on $\gammamsone$ and $\estart$.
The general behavior does not depend on the starting energy $\estart$.
This is compatible with our previous observation that the strain value at the stress overshoot does not depend on $\estart$.
When studying the dependence on $\gammamsone$, we find that for $\gammamsone < \gammac$ the peak position corresponds to the ensemble averaged peak position.
This shows again that these transitions have no effect on the yield position and are irrelevant.
Indeed, it can generally be shown (see Appendix~\ref{sec:firstms} FIG.~\ref{fig:smsubstressreg1}) that the results in the regime of $\gammamsone > \gammac$ are independent of whether or not MS transitions were already present for $\gammamsone < \gammac$.
For $\gammac < \gammamsone < 0.15$, the peak follows closely the MS transition.
Thus, we may conclude that the peak is caused by the MS transition.
In a more detailed view it turns out that $\gammamsone$ is always larger than $\gammapeak$.
This is at least partly due to the averaging of individual highly non-symmetric stress-strain curves.
Finally, for $\gammamsone > 0.15$ the overshoot peak hardly changes with $\gammamsone$ and starts to become unrelated to the first MS transition.
In this case the stress relaxation of the IS transitions during the first MS effectively create a stress overshoot; see Appendix~\ref{sec:firstms} FIG.~\ref{fig:smsubstressreg3}.
Since, by definition, trajectories are reversible by AQS shear with $\gamma < \gammamsone$, the overshoot peak is visible although at this strain value the system is still related to its initial MS.

\section{Reversibility and AQS Cyclic Shear}
A lot of information about the impact of shearing on the dynamics within the  PEL can be gained from cyclic shear simulations with amplitude $\gammamax$.
A key concept, discussed in literature \cite{Fiocco2013,Kawasaki2016}, is the concept of limit cycles.
In order to explore the properties of small systems (again $N=130$) with respect to limit cycles we performed AQS cyclic shear simulations in dependence of $\gammamax$ for 128 different initial conditions.
Here we always start from $\estart \approx -4.60$.

      \begin{figure}[tbp]
    \centering
   \includegraphics[width=0.51\textwidth]{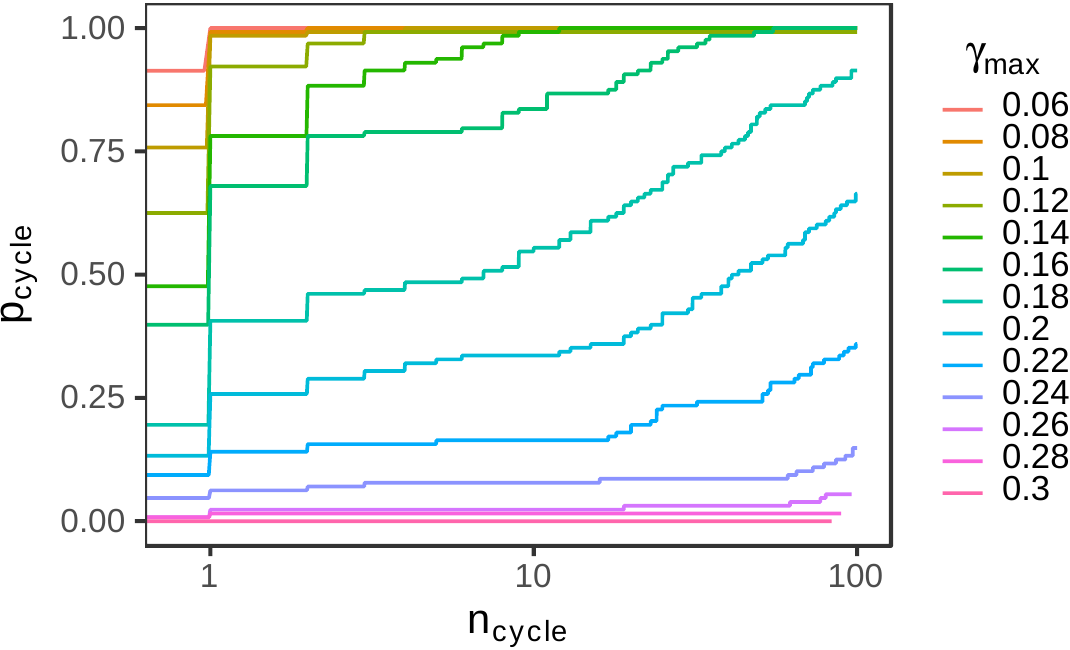}
    \caption{\label{fig:aqs_pcycle}
Probability that after $n_{cycle}$ cycles the system has visited a limit cycle.
    }
\end{figure}

In FIG.~\ref{fig:aqs_pcycle} we show the probability that for given $\gammamax$ the system has entered a limit cycle after $n_{cycle}$ cycles.
Data are shown for up to 100 cycles.
As expected, for small values of $\gammamax$ already after a few cycles the system is in a limit cycle with basically 100\% probability whereas in the opposite limit of large values only a tiny fraction has managed to reach a limit cycle for $n_{cycle} = 100$.

\begin{figure*}
	\subfloat[]{\label{fig:aqs_firststart}
		\includegraphics[width=0.45\textwidth]{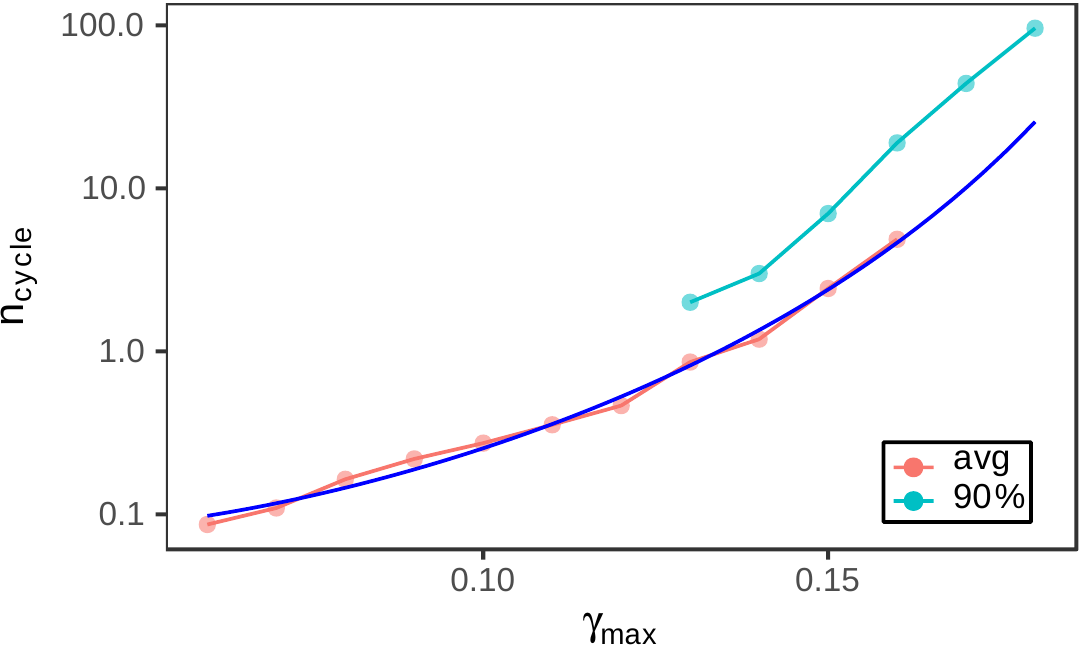}
    }
    \subfloat[]{\label{fig:aqs_pcycle1}
        \includegraphics[width=0.45\textwidth]{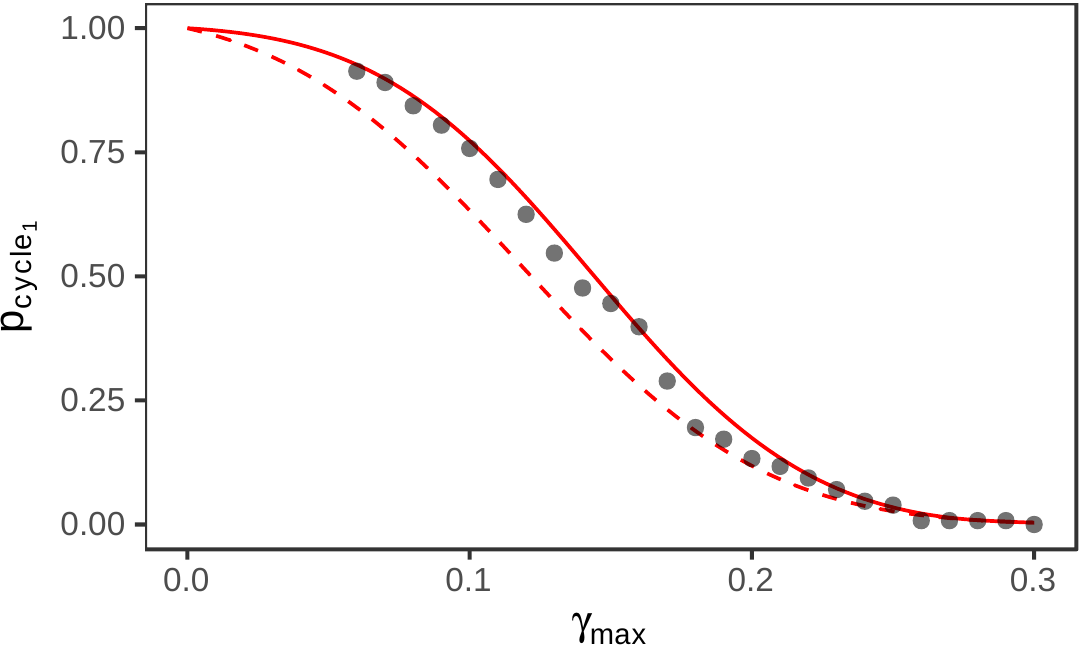}
    }
    \caption{\label{fig:aqs}%
(a)  Red points:  Average value of $n_{cycle}$, where a
limit cycle is entered in cyclic AQS simulations at $N = 130$ as a
function of $\gammamax$.  The dark blue line denotes a power law
fit of the form $\gammastart = c \cdot (\gammanl - \gammamax)^\delta$ with
the fit parameters $c = 1.44 \cdot 10^{-6}$, $\delta = -5.49$ and $\gammanl = 0.231$.
Blue points: The value of $n_{cycle}$ where 90\% of all
simulations have entered a limit cycle.  (b) Probability to be
already in a limit cycle at the start of a cyclic AQS simulation
at $N = 130$. The solid red line represents an approximation using
MS statistics (see text). For comparison the dashed red line shows
$P(\gammams > \gammamax)$.
        }
    \end{figure*}

For a more detailed analysis of this data we first ask how long it takes to reach a limit cycle.
Results for the average number of cycles are shown in FIG.~\ref{fig:aqs_firststart}.
One can clearly see that the average accumulated strain strongly increases
with increasing value of $\gammamax$. A power law fit diverges at
$\gammanl = 0.231$. This result can be directly compared with the
results from \cite{Kawasaki2016} where a divergence is seen as
well, albeit at much smaller values of $\gammamax$. Since the
critical amplitude strongly depends on system size \cite{Kawasaki2016,
Fiocco2013} it is not surprising that we obtain values larger than 0.20
for $\gammanl$. We would like to stress that the onset of the divergence in the analyzed range of $\gammamax$-values is quite weak. Furthermore, we also
show when 90\% of all simulations have entered a limit cycle.
Comparison with the average value indicates that the
heterogeneity becomes larger for larger $\gammamax$.

\begin{figure}[tbp]
    \centering
	\includegraphics[width=0.5\textwidth]{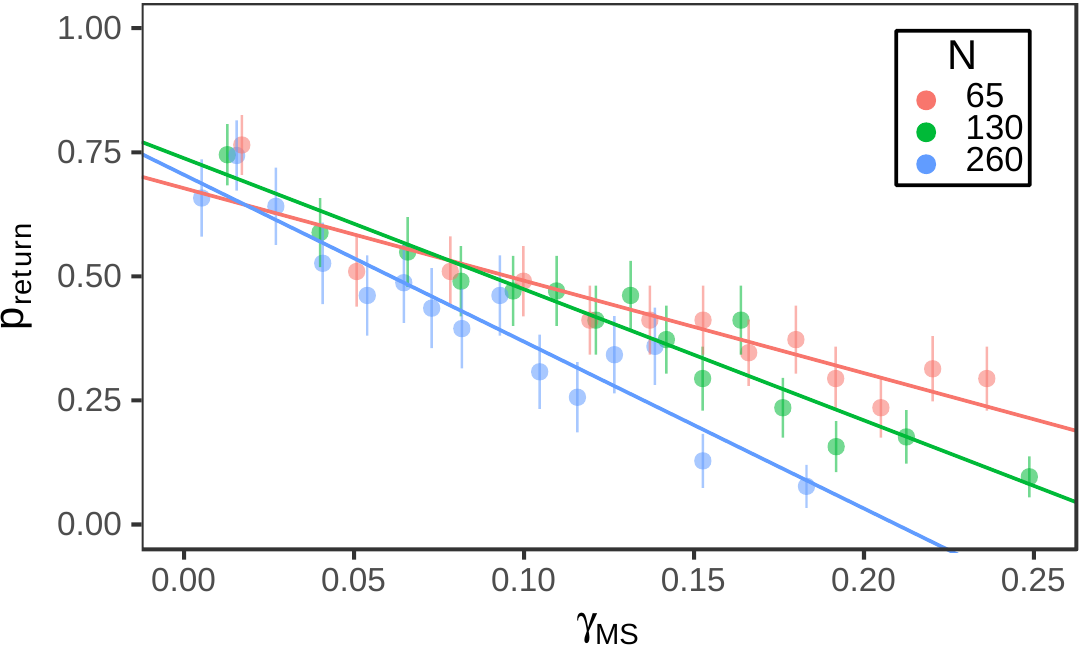}
    \caption{
            \label{fig:preturn}
Probability to visit the starting MS by AQS shear in the reverse
direction (up to $\gamma = -0.3$) after the first MS transition in
non-cyclic simulations. The solid lines represent linear fits to
the data and for $N = 130$, it has a slope of $-2.64(20)$ and an
intercept of $0.74(3)$. By extrapolation, $\preturn$ is zero for
$\gammams \ge 0.279(10)$ at $N = 130$.
    }
\end{figure}

Another aspect from FIG.~\ref{fig:aqs_pcycle} deals with the
probability $\pcycleone$ that already in the initial configuration
the system has entered a limit cycle. The dependence of this
probability on $\gammamax$ is shown in FIG.~\ref{fig:aqs_pcycle1}. For a closer understanding of $\pcycleone$ the concept of MS turns
out to be very helpful. Naturally, if there is no MS transition up
to  $\gamma = \gammamax$ in both shear directions for a single
trajectory, then, under cyclic AQS shear, a limit cycle will form
at all amplitudes smaller than $\gammamax$. For higher amplitudes
the situation will become more difficult, as MS transitions may be
reversible or not. The probability for the reversibility of a MS
transition $\preturn$ with respect to its position $\gammams$ is
given in FIG.~\ref{fig:preturn}. Even for very early MS
transitions (e.g., $\gammamax  = 0.1$) there is a high chance $\approx 25\%$ to be
irreversible. In the limit of high $\gammamax$, we find that the
upper limit of reversibility is at $\gammams = \gammaul \approx
0.28$. As the return probability for higher $\ems$ is lower and
the MS position $\gammams$ would be shifted towards the current
shear direction by an MS transition, $\gammaul$ should form an
upper limit on the overall reversibility. $\gamma_{\textrm{ul}}$
compares well with our results from cyclic AQS simulations, as we
find less than 1\% limit cycles in simulations with $\gammamax =
0.29$ and none with $\gammamax = 0.3$. The strong system size dependence in
$\gammaul$, we find for our small systems, is expected
in analogy to the behavior of $\gammanl$
\cite{Fiocco2013,Kawasaki2016}.

In close relation to $\preturn$ is the probability to find a
limit cycle directly at the start of the simulation $\pcycleone$;
see FIG.~\ref{fig:aqs_pcycle1}.
To find a limit cycle directly at the start, there must be either no MS transition or a reversible one in both shear directions.
We approximate this probability with the expression:
    \begin{align}
        & \pcycleone(\gammamax) \approx \\
        & \qquad P(\gammamsone > \gammamax) + P(\gammamsone \le \gammamax) \preturn(\gammamax) \cdot \nonumber \\
        & \qquad (P(\gammamsone > \gammamax) + P(\gammamsone \le \gammamax) \preturn(\gammamax)) \nonumber
    \end{align}
Although the first term dominates $\pcycleone(\gammamax)$, a very close description can be only achieved by using the full expression; see FIG.~\ref{fig:aqs_pcycle1}.
We rationalize the complete expression as follows:
If there is no MS transition until $\gammamax$, we also expect no MS transition in the reverse direction. In the opposite case, i.e. $\gammamsone < \gammamax$,
 $\pcycleone$ is determined by the reversibility of the previous MS transitions.
Because there may be multiple MS transitions before reaching $\gammamax$, we assume that the last transition is close to $\gammamax$.
Furthermore, if the last transition is reversible, the others before must have also been reversible and $\preturn$ for the last transition should have the same value, as if it was the first transition.
Therefore, we can use $\preturn(\gammamax)$ instead of a more complicated term.
If there was an MS transition on the forward direction, we assume that there may also be MS transitions on the backward direction, which may also be irreversible, forcing us to evaluate $P(\gammams > \gammamax)$ again.
For the evaluation of the $\preturn$ in the reverse direction, the same argumentation as for the forward direction holds.

\begin{figure*}
    \subfloat[]{
       	\label{fig:cycleenergy_ncycle}
		\includegraphics[width=0.45\textwidth]{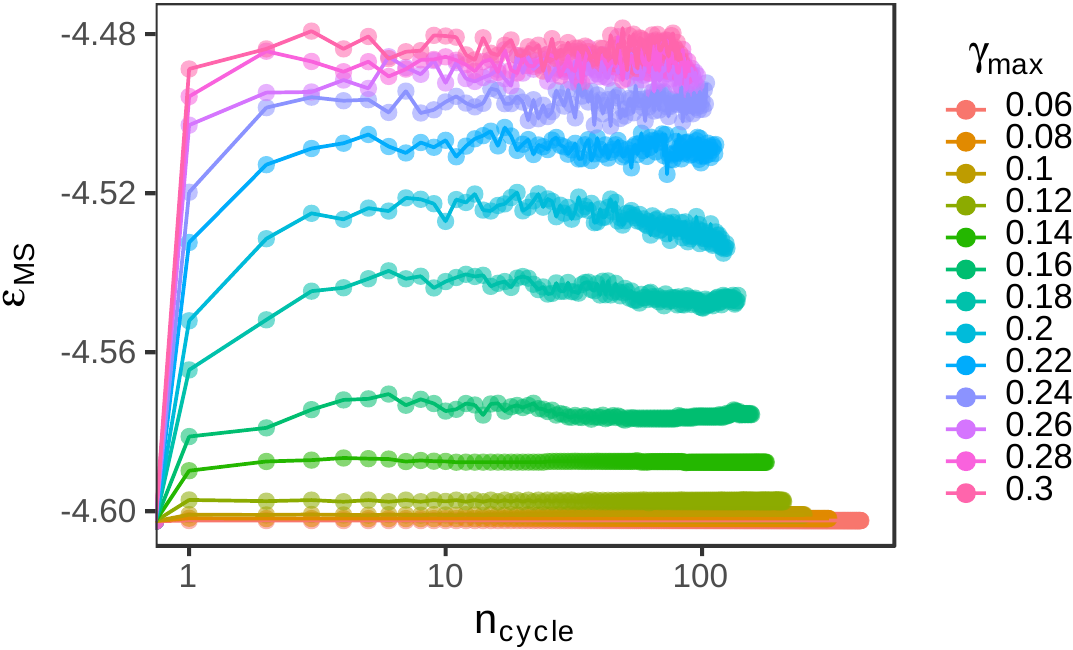}
    }
    \subfloat[]{
        \label{fig:cycleenergy_gammamax}
        \includegraphics[width=0.45\textwidth]{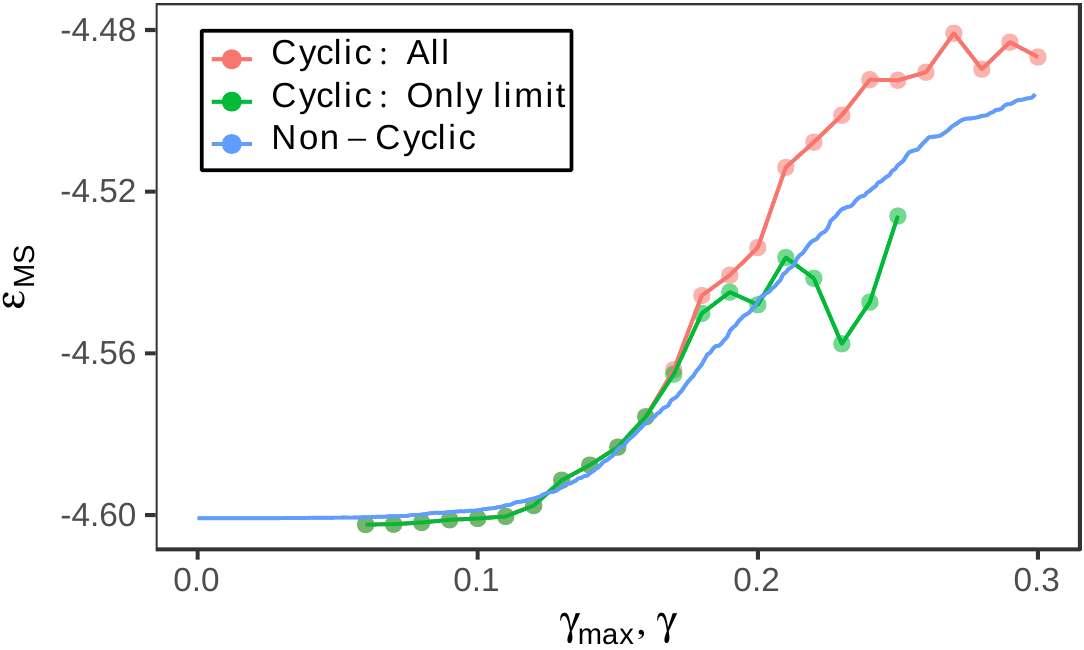}
    }
    \caption{\label{fig:cycke}%
		(a) The average MS energy after $n_{cycle}$ cycles for different values of $\gammamax$.
		(b) The final MS energy after 100 AQS cyclic shear simulations together with the final MS energy under the condition that the system is in a limit cycle after 100 cycles.
		Additionally the MS energy from FIG.~\ref{fig:energyens} is included.
    }
\end{figure*}

Finally, it is instructive to connect the observation of limit cycles with the underlying PEL properties and to compare it with the corresponding results in \cite{Fiocco2013}.
As shown in FIG.~\ref{fig:cycleenergy_ncycle}, already after a few cycles the limiting energy is reached.
One roughly finds that around $\gammamax = 0.2$ it takes the longest time (approx. 7 cycles) to reach the maximum energy.
In contrast, in \cite{Fiocco2013} it took for the $N=4000$ system  more than approx. 50 cycles to reach the final energy value for $\gammamax = 0.08$.
This shows for the large system how the elastic interaction among different subunits gives rise to a slow increase of energy in contrast to the results for small systems.

Furthermore, it turns out that the final energy, shown in FIG.~\ref{fig:cycleenergy_gammamax} as a function of $\gammamax$, closely resembles the MS energies, already shown in FIG.~\ref{fig:energyens}. Here we distinguish whether or not after 100 cycles the system has reached a limit cycle or not.  We can distinguish three $\gamma$-regimes. For $\gammamax \le 0.16$ the energy at the turning point $\gamma = \gammamax$ in the first cycle is the same as the energy in the 100th cycle. Among others this implies that the approx. 60\% trajectories which for $\gammamax = 0.16$ did not start in a limit cycle (see FIG.~\ref{fig:aqs_pcycle1}) hardly increase their MS energy upon reaching a limit cycle in the course of oscillatory shearing. For $0.17 \le \gammamax \le 0.19$ by far the most trajectories still end up in a limit cycle so that the final MS energy after 100 cycles hardly changes upon restriction to the subset of trajectories with a limit cycle. However, the MS energies are now significantly higher than in the first cycle. Finally, for $\gammamax \ge 0.20$  the MS energies remain constant within statistical uncertainty, only taking into account the trajectories in a limit cycle.
This suggests that the energy of approx. -4.55 is a kind of crossover energy in the PEL.
Once the system has significantly crossed this energy, the chance to enter a limit cycle approaches zero, independent of the applied amplitude $\gammamax$. Remarkably, this energy cutoff is lower than the average energy at the turning point of the first cycle (for $\gammamax \ge 0.22)$. Thus, already after one cycle most of the trajectories have left the phase space where a limit cycle is still possible.

We would like to point out that the increase of the MS energy as a function of $\gammamax$ (taking the average over all trajectories after 100 cycles) is very continuous. In contrast, for the $N=4000$-particle system in \cite{Fiocco2013} one finds a discontinuous behavior when crossing the value of $\gammamax$  where limit cycles start to become relevant (between 0.08 and 0.09 for the large system).
This is consistent with our previous observation in FIG.~\ref{fig:cycleenergy_ncycle} that for all values of $\gammamax$ the final energy is reached after a few cycles so that there is not much space for a significant increase of energy for trajectories which do not enter a limit cycle so early (or at all). Thus, also the strong increase of energy with the number of cycles in \cite{Fiocco2013} can be related to the elastic coupling of the different subsystems.

\section{Discussion and Summary}
The framework of the PEL $V(\x)$ can be used for a systematic analysis of quiescent glass-forming systems.
By considering the extended PEL $V(\x,\gamma)$ and incorporating the new concept of minimized structures (MS), one can also capture the underlying properties of yielding.
Thus, for the case of shearing the MS generalize the concept of IS by taking into account an additional variable for minimization. 
Furthermore, for a system just containing $N = 130$ particles it is possible to analyse more elementary plastic events as opposed to larger systems where significant coupling effects due to the elastic interaction among different regions of the glass are present as well \cite{Leishangthem2017}.

We could identify two strain values $\gammac \approx 0.09$ and $\gammacc \approx 0.20$ where a specific crossover behavior is observed.
For $ \gamma \approx \gammac$ the MS transitions become relevant in the sense that on average observables such as the energy start to change significantly.
As a consequence, neither for simple shearing nor for cyclic shearing any significant effects are expected for $\gamma < \gammac$.
The second crossover strain $\gammacc$ can be observed from AQS cyclic shear simulations with amplitude $\gammamax$.
Around $\gamma \approx \gammacc$ the number of trajectories, ending up in a limit cycle, strongly decreases because the natural MS energy upon shearing starts to become larger than the MS energy which is still compatible with the presence of limit cycles.
Furthermore, the number of cycles, required to reach the stationary state, is a maximum for $\gammamax \approx \gammacc$. Thus, the properties of $\gammacc$ are directly comparable with those of $\gammanl$, as previously reported for large systems. Therefore it is tempting to identify $\gammacc$ with the value of $\gammanl$ in the limit of small systems. 

Furthermore, when studying the first relevant MS transition, it turns out that the overshoot in the stress-strain curve is governed by the first MS transition as long as that transition occurs for $\gamma \le 0.15$.
Thus, we can relate the observation of yielding to properties of the extended PEL $V(\x,\gamma)$.
However, if there is no MS transition for $\gamma \le 0.15$, the overshoot maximum is governed by earlier IS transitions which can also give rise to stress release.
In general, the overshoot maximum in this limit is much broader than the overall average.

The introduction of MS has another advantage.
In quiescent equilibrium simulations at low temperatures, metabasins can be constructed directly from the trajectory of inherent structures by detecting transitions to previously visited structures using the algorithm detailed in \cite{Doliwa2003c}.
For the sheared system the IS changes with strain so that the metabasin construction would break down, even for very low shear rates and high temperatures where the physics of the quiescent case should still be relevant.
For this limit it is very helpful that for every IS there exists a unique minimized structure.
Then metabasins can be constructed on the basis of MS rather than IS.
Furthermore, this replacement is even beneficial for the quiescent case, as it may associate several IS with the same MB by using the structural connection via the MS without the need of backward jumps.
As metabasins can be used to explain $\taualpha$ in terms of a continuous time random walk \cite{Rubner2008} and diffusion \cite{Doliwa2003b} for the quiescent system, we anticipate that our extended definition can provide new insights into the transition from shear to temperature dominated behavior of the sheared system.

Is it possible to rationalize the critical value of $\gammanl \approx 0.07-0.08$, observed for oscillatory shear simulations of large systems \cite{Fiocco2013,Leishangthem2017}, with the insight from the simulations of small systems? Remarkably, this critical value is close to the crossover strain value $\gammac \approx 0.09$ from which on relevant MS transitions do occur. Here is a suggestion for a simple scenario:  Our basic hypothesis is the possibility to view a large system  as a superposition of small systems, displaying a strong elastic interactions among each other. This interaction gives rise to a redistribution of the stress.  The elementary plastic events are localized in the subsystems. For $\gammamax < \gammac$ the individual subsystems only display irrelevant MS transitions. Naturally, due to the elastic coupling the local (small) stress relaxation, MS transitions may be triggered in nearby subsystems since the local PEL has changed. Due to the smallness of irrelevant MS transitions the probability of an induced MS transition, however, is small. Furthermore, due to the small overall value of $\gammamax$ these triggered MS transitions are likely also to be irrelevant. As a consequence of both effects, the avalanche-type behavior would remain localized \cite{Leishangthem2017}. Furthermore, also the resulting increase of the energy until reaching a limit cycle remains small \cite{Fiocco2013}.  The scenario is very different for $\gammamax > \gammac$. Individual subsystems without elastic interaction would still enter a limit cycle (as long as $\gammamax < \gammacc \approx 0.2)$. However, due to the interaction and the presence of relevant MS transitions the individual energy/stress/strain variations upon plastic events are sufficiently large to induce with a much higher probability a similar (and also relevant) MS transition in a nearby subsystem. This suppresses the possibility of limit cycles for larger systems.  Now these coupling effects do not fade away but induce transitions in the whole system. This results in a strong increase of the energy with the number of cycles \cite{Fiocco2013} as well as in the emergence of a system-spanning avalanche behavior \cite{Leishangthem2017}. This scenario is also compatible with the increasing smearing out of the yielding transition (beyond the shift of the crossover strain value)  when decreasing the system size \cite{Jaiswal2016}. Interestingly, already in the range from $N=65$ to $N=260$ particles the system-size dependence is visible for $\gammamax > \gammac$. Please note that for the whole discussion it was implied in agreement with the results of FIG.~\ref{fig:gammams_enextms_IE} that the crossover between irrelevant and relevant MS transitions does not depend on the present MS energy. This simplifies the above discussion because an energy dependence of $\gammac$ would have influenced the coupling effects during the continuous increase of the energy for $\gammamax > \gammac$ (low energy limit). Finally, this scenario is also compatible with the strong finite size effects, observed for the value of $\gammanl$. Efficient coupling mechanisms have to be present to reduce the value from $\gammanl \ge 0.2$ for $N=130$ to 0.07-0.08 in the limit of large system sizes and thus to avoid the presence of limit cycles for, e.g., $\gammamax = 0.1$ in that limit. This requires a large number of interacting subsystems in order boost the resulting coupling effect. Also in other fields of physics the resulting effective fluctuations often increase with system size. For the example of a ferromagnet this gives rise, e.g.,  to finite-size effects of the phase transition temperature.

In principle, the validity of this type of scenario is testable in the framework of elasto-plastic models \cite{Martens2012} if the actual properties of the subsystems are taken into account. It might be also interesting to analyse the relation between $\gammapeak$ and $\gammanl$ in this class of models.  We hope that the present work about the detailed characterization of the plastic events and yielding properties of glasses with small system size will prompt work along this line. In any event, in analogy to the quiescent case important information can be hidden in the small-system world \cite{Heuer2008,Rehwald2012} to better understand the relevant macroscopic limit.

\thanks{We thank the DFG via FOR 1394 for financial support and J.L. Barrat, K. Martens, M. Maiti, and L. Smith for helpful discussions.}

\appendix
\section{\label{app:straindef}Overview of strain definitions}
\begin{table}[h!]
	\begin{ruledtabular}
		\begin{tabular}{c|l}
		    \hspace{2em}$\gammapeak$\hspace{2em} & Position of the stress overshoot maximum. \\
		    $\gammamsone$   & Position of the first MS transition in a single trajectory. \\
		    $\gammac$		& Average position of the first MS transition, where a crossover is seen in central observables. \\
		    $\gammaul$		& Extrapolated maximum strain, beyond which there are no reversible MS transitions. \\
		    $\gammamax$		& Strain amplitude in oscillatory shear. \\
		    $\gammanl$		& Average strain amplitude in oscillatory shear, beyond which no limit cycles can be found. \\
		    $\gammacc$		& Strain amplitude in oscillatory shear, where the MS energy in limit cycles reaches its maximum. \\
		\end{tabular}
	\end{ruledtabular}
\end{table}

\section{\label{app:jumpdetection}Coincidence Probability of different Transition Types}
\begin{table}[h!]
	\begin{ruledtabular}
		\begin{tabular}{c|cccc}
			$P(X | Y)$      & $d$   & $\dmin$   & $\epsilon$    & $\sigma$ \\
			\hline
			$d$             &       & 0.994     & 0.75          & 0.86 \\
			$\dmin$         & 0.90  &           & 0.70          & 0.81 \\
			$\epsilon$      & 0.92  & 0.94      &               & 0.84 \\
			$\sigma$        & 0.90  & 0.94      & 0.73          & \\
		\end{tabular}
	\end{ruledtabular}
	\caption{\label{tab:pjumpstats}
		Probabilities that a transition of type $X$ is detected at the position of a transition of type $Y$ in the IS trajectory.
		$X$ is given in horizontal and $Y$ in vertical direction.
		($d$ : euclidean distance of all particles, $\dmin$ : non-affine displacement of neighboring particles, $\epsilon$ : energy, $\sigma$ : stress).
	}
\end{table}

\section{\label{app:compisms}Comparison of IS and MS transitions}
\begin{figure}[tbp]
	\includegraphics[width=0.5\textwidth]{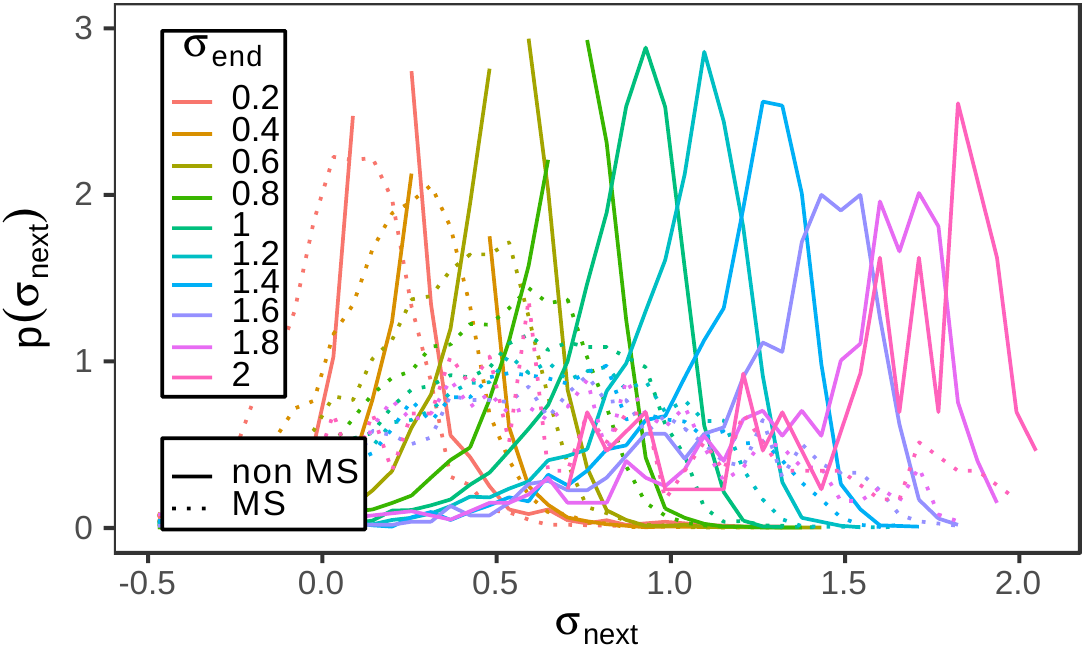}
	\caption{\label{fig:dminjumpstats}
		Probability density $p(\sigmanext | \sigmaend)$.
		The stress before the MS/non-MS transition is called $\sigmaend$ and the stress after the transition $\sigmanext$.
		To calculate the probabilities, we used all transitions with $\gamma \le 1.0$, including both the overshoot region and the flow regime.
	}
\end{figure}
MS transitions are basically a subset of the IS transitions.
The main difference is an additional change in the MS energy value at the transition.
This makes this transition irreversible by straining in the opposite direction.
However, there is an additional difference, which can be seen in FIG.~\ref{fig:dminjumpstats}.
For non-MS jumps the distribution of stress values $\sigmanext$ after the transition, has a very strong peak near the stress before the transition $\sigmaend$.
Compared to the stresses at MS transitions the width of the distribution is small.
Starting at high $\sigmaend$, there is only a low probability reaching stresses at or below the flow regime ($\sigma \le 0.7$).
So non-MS jumps are unlikely to show large stress drops, which explains the long plateau in FIG.~\ref{fig:smsubstressreg3}.
In contrast, MS jumps show a broad distribution of $\sigmanext$ centered around the mean stress of the flow regime, hinting at a large redistribution of stresses in the system, basically losing much of the previous stress history.
For $\sigmaend$ smaller than the flow stress, this distinction starts to break down, as both distributions show more overlap.

\begin{figure}[tbp]
	\includegraphics[width=0.45\textwidth]{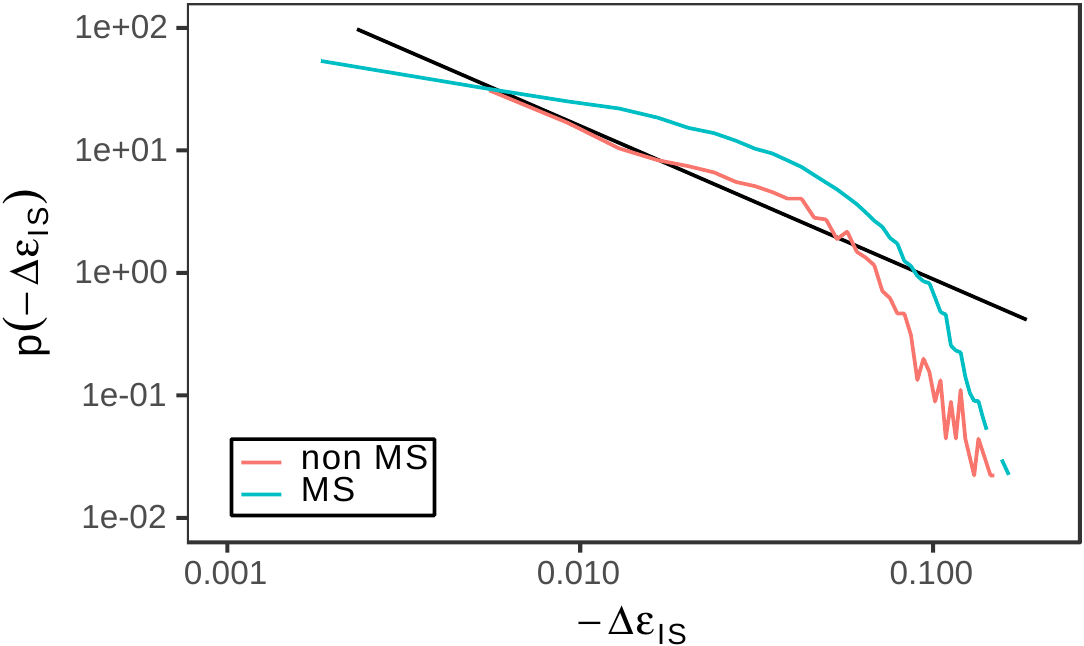}
	\caption{\label{fig:energydrops}%
		Distribution of energy drops at MS and non-MS transitions.
		The straight black line reflects a power-law behavior with a slope of $-\frac{5}{4}$.
	}
\end{figure}
A similar difference can also be seen in the distribution of energy drops, where we find higher drops at MS transitions than at IS transitions.
The characteristic power-law scaling with a slope of $-\frac{5}{4}$ is thereby only seen in the energy drops at IS transitions.

\section{\label{app:firstms}First minimized structure}
\begin{figure}[tbp]
	\centering
	\includegraphics[width=0.45\textwidth]{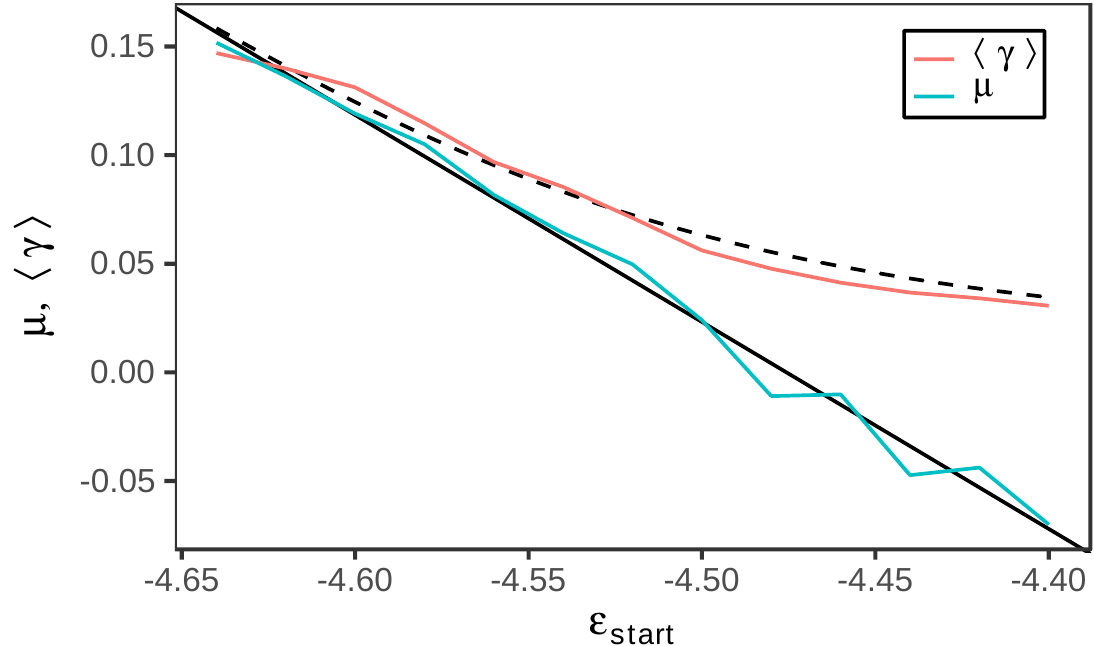}
	\caption{\label{fig:gammams_scaling}
		Parameterization of the $\gammamsone$ distribution, which has a cut off Gaussian shape.
		The blue line shows the fit parameter $\mu$.
		The black straight line represents a fit to these values with a slope of $-0.953(28)$ and an intercept of $-4.26(13)$.
		The red line shows the average value of $\gammams$, calculated directly from the distribution.
		From the linear fit of $\mu$, we also calculated the average $\gammams$, which is shown as the black dashed line.
		For all cut off Gaussian distributions, we used a fixed $\sigma = 0.0675$, which we determined as the mean fit parameter for the lowest $\estart$.
	}
\end{figure}
The distribution of the position of the first MS transition has a Gaussian shape with a cutoff at $\gammamsone = 0$, as shown in FIG.~\ref{fig:gammams_distribution}.
The position of the maximum of the Gaussian $\mu$ decreases linearly, as shown in FIG.~\ref{fig:gammams_scaling}.
As it is difficult to determine the width at high $\estart$ due to the cutoff, we used a fixed with of $\sigma = 0.0675$.
This value was derived by Gaussian fits with variable width over the distributions with the four lowest $\estart$.

\begin{figure}[tbp]
	\includegraphics[width=0.45\textwidth]{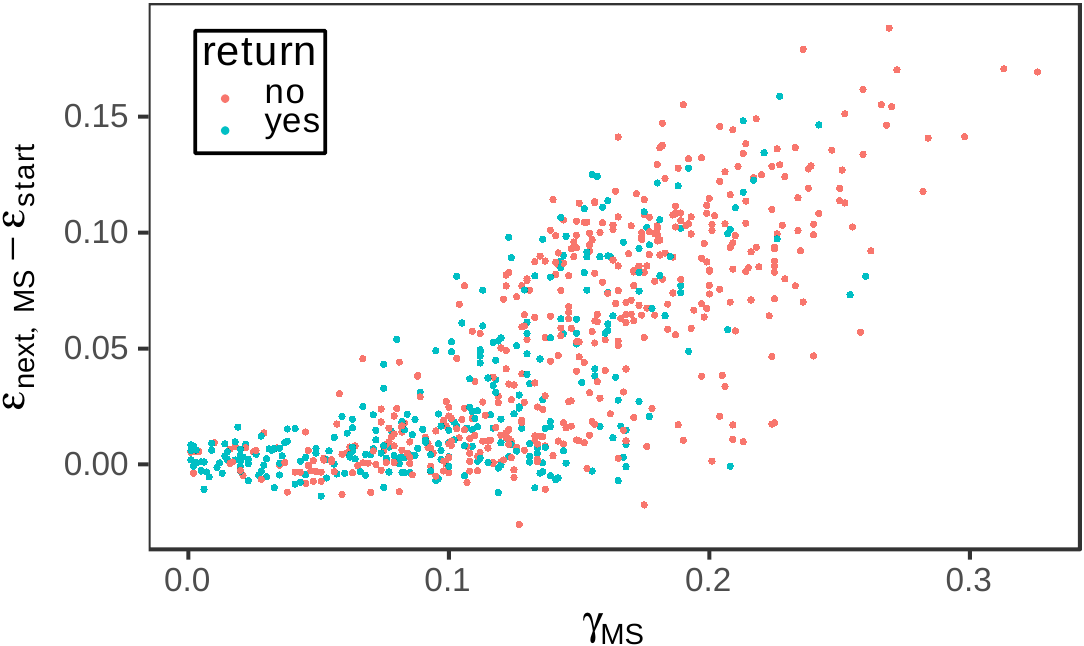}
	\caption{\label{fig:notsignificant}%
		Scatter plot of the position of the first MS transitions.
		The color shows, if the starting structure can be found be shearing in the reverse direction until $\gamma = -0.3$ immediately after the MS transition.
	}
\end{figure}
In Section~\ref{sec:firstms}, we found a crossover in the average energy change at a given $\gammams$.
As can be seen from FIG.~\ref{fig:notsignificant}, there is no sharp transition when looking at single trajectories.
Rather, we see two overlapping point clouds, characterizing the two different kinds of processes.
Furthermore, we find that there is no correlation between the two processes and reversibility, showing that also large changes in energy can be reversible by shearing in the backward direction.

\begin{figure*}[tbp]
	\subfloat[]{
		\label{fig:smsubstressreg1}
		\includegraphics[width=0.33\textwidth]{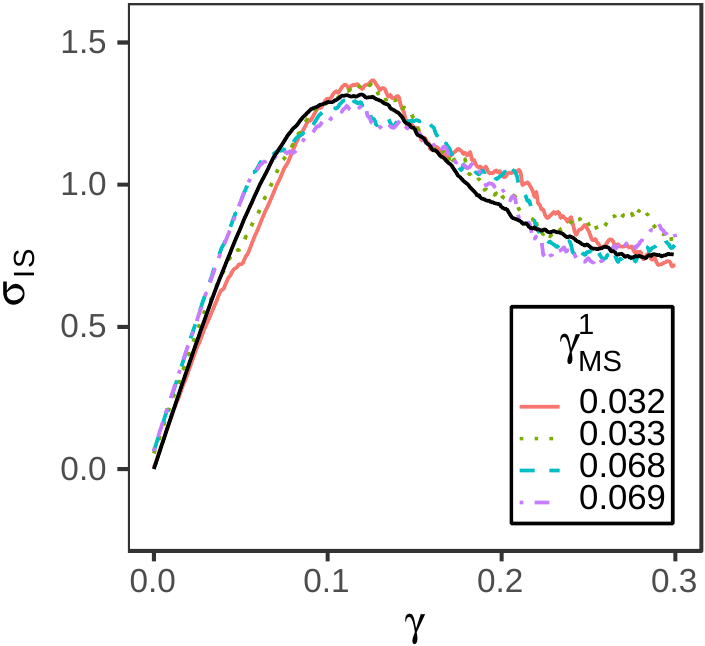}
	}
	\subfloat[]{
		\label{fig:smsubstressreg2}
		\includegraphics[width=0.33\textwidth]{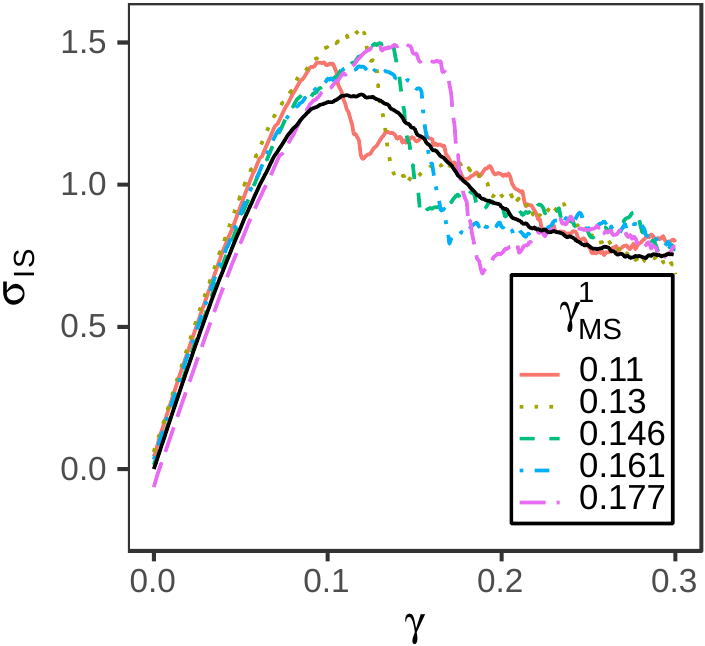}
	}
	\subfloat[]{
		\label{fig:smsubstressreg3}
		\includegraphics[width=0.33\textwidth]{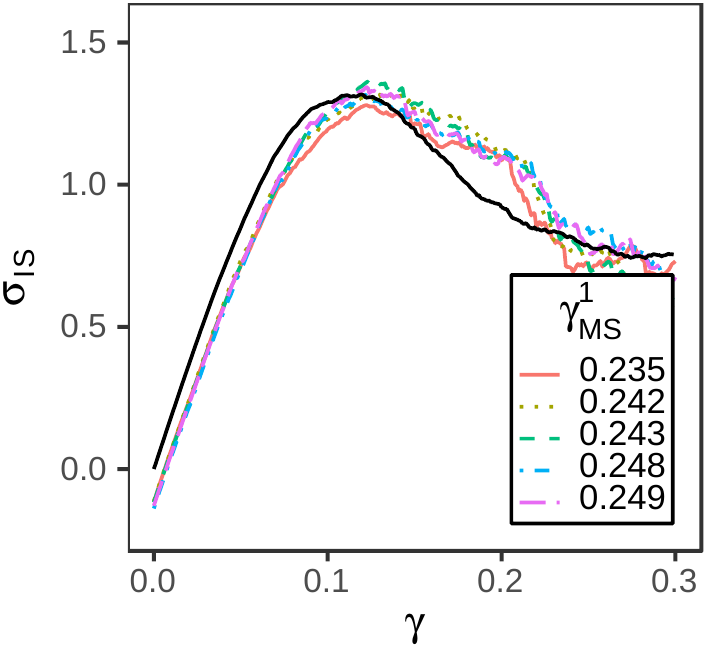}
	}
	\caption{\label{fig:smsubstress}
		a) Stress of sub-ensembles with a given strain $\gammams < 0.09$. Show are trajectories for temperatures $T \in \{1 \cdot 10^{-4},5 \cdot 10^{-4}\}$.
		b) The same for $\gammams \in (0.09,0.2)$ and $T = 10^{-4}$.
		c) For $\gammams > 0.21$ we plot all available values for $T \in [10^{-4},10^{-2}]$.
	}
\end{figure*}
In Section~\ref{sec:firstms}, we have shown the relation between the overshoot maximum and the first MS transition.
We now show the respective stress strain curves for the respective subensembles in FIG.~\ref{fig:smsubstress}.
The values for $\gammamsone \le 0.07$ show, that there is no correlation between the peak position and the first MS jump in this regime.
In FIG.~\ref{fig:smsubstressreg1}, some sub-ensembles from trajectories in this region are shown.
One can see, that for each sub-ensemble, the stress resembles closely the ensemble average over all trajectories.
This means, that these transitions have no special contribution to the stress overshoot.

In the region $0.09 \le \gammamsone \le 0.2$ (FIG.~\ref{fig:smsubstressreg3}) we see that there is a large stress drop in each sub-ensemble.
The stress drop occurs sharply at or slightly after the peak position.
The shift can be explained by small IS transitions, which may happen before the MS transition.

The trajectories for the region $\gammamsone > 0.2$ are shown in FIG.~\ref{fig:smsubstressreg3}.
We observe that all trajectories start with a slightly negative stress.
Since we simulate a small system with fixed boundary conditions the stress after minimizing the potential energy may be non-zero, also in the fully equilibrated system.
This leads to a slight $\gamma$ offset to the ensemble average until the point where we see a plateau.
This plateau is almost as high as the ensemble average, so it is built by small stress drops, which cancel out in the sub-ensemble.
The stress is kept high because of the high energy difference between the low energetic MS energy and the current already high IS energy after the small stress drops.
The plateau then ends with a sudden drop in stress, which is caused by a large stress released together with the MS transition.

\FloatBarrier

\bibliographystyle{apsrev4-1}
%

\vfill\eject

\end{document}